\documentclass[twocolumn,showpacs,superscriptaddress,groupedaddress]{revtex4-1}  % for review and submission
\usepackage{lipsum}
\usepackage{graphicx}  % needed for figures
\usepackage{dcolumn}   % needed for some tables
\usepackage{bm}        % for math
\usepackage{amssymb}   % for math

\usepackage{amsmath}
\usepackage{algcompatible}
\usepackage{newfloat}
\usepackage{color}
\usepackage{bbm}
\usepackage{hyperref}

\usepackage{blindtext}
\usepackage{chngcntr}
\counterwithin*{section}{part}

\DeclareFloatingEnvironment[
    fileext=loa,
    listname=List of Algorithms,
    name=ALGORITHM,
    placement=tbhp,
]{algorithm}

\begin{document}
\title{A parallelizable sampling method for parameter inference of large biochemical reaction models}
\author{Jan Mikelson}
\author{Mustafa Khammash}
%%\input author_list.tex       % D0 authors (remove the first 3 lines
%%                             % of this file prior to submission, they
%%                             % contain a time stamp for the authorlist)
%%                             % (includes institutions and visitors)
\date{\today}

\begin{abstract}
The development of mechanistic models of biological systems is a central part of Systems Biology. One major task in developing these models is the inference of the correct model parameters. Due to the size of most realistic models and their possibly complex dynamical behaviour one must usually rely on sample based methods. In this paper we present a novel algorithm that reliably estimates model parameters for deterministic as well as stochastic models from trajectory data. Our algorithm samples iteratively independent particles from the level sets of the likelihood and recovers the posterior from these level sets. The presented approach is easily parallelizable and, by utilizing density estimation through Dirichlet Process Gaussian Mixture Models, can deal with high dimensional parameter spaces. We illustrate that our algorithm is applicable to large, realistic deterministic and stochastic models and succeeds in inferring the correct posterior from a given number of observed trajectories. This algorithm presents a novel, computationally feasible approach to identify parameters of large biochemical reaction models based on sample path data.
\end{abstract}

\maketitle

\part*{}
\section{Introduction}

%{\it \color{red}Biomolecular models and their importance}
\par{The accurate modelling and simulation of biological processes such as gene expression or signalling has gained a lot of interest over the last years and a large body of literature  discussing various types of models, their inference and means of simulation has emerged. The main purpose of these models is to reflect and faithfully reproduce observed biological dynamics, while giving an insight into the underlying bio-molecular mechanisms. 
}

%{\it \color{red}Importance of parameter inference (Relevance)}
\par{One important aspect in the design of these models is the determination of the model parameters. Often there exists a mechanistic model of the cellular processes but parameters (e.g. reaction rates or initial molecule concentrations) are not known. Since the same network topology may result in different behaviour depending on the chosen parameters (\cite{ingram2006network}), this presents a major problem for model inference and explains the need for parameter estimation techniques.}

%{\it \color{red} Deterministic and stochastic models} 
\par{The models used in Systems Biology can be coarsely classified into two groups: deterministic and stochastic models. Deterministic models usually rely on ordinary differential equations which, given the parameters and initial conditions, can describe the development of the biological system in a deterministic manner. However, many cellular processes like gene expression are subject to stochastic fluctuations. Traditionally such models have been treated deterministically, relying on the large copy numbers of the involved molecules to take care of any stochastic effects (see \cite{karlebach2008modelling} for a review). It has been indicated though that the randomness of these systems may play an important role  (\cite{elowitz2002stochastic}, \cite{mcadams1997stochastic}, \cite{ozbudak2002regulation}), stirring an increasing interest in stochastic models over the last years (\cite{lillacci2013signal}, \cite{zechner2014scalable}, \cite{neuert2013systematic}, \cite{hilfinger2011separating}, \cite{boys2008bayesian}). Such stochastic models are usually described in the framework of stochastic chemical reaction networks that can be simulated using the Stochastic Simulation Algorithm (SSA) (\cite{gillespie1977exact}).
}

%{\it \color{red}The data and lack of inference techniques } 
\par{The inference of the model parameters is done from mainly two kinds of biological data. First, distribution data, which is data that reflects a whole population of cells at a particular time point and shows the distribution of the measured output in the entire population. The other is trajectory data, which focuses on single cells and monitors the behaviour throughout time. In the past years the availability of trajectory data has drastically increased, providing detailed information of the development of single cells throughout time. Unfortunately, while there has been plenty of research on parameter estimation for deterministic systems (\cite{toni2009approximate}, \cite{putter2002bayesian}, \cite{timmer2004modeling}, \cite{bortz2006model}) and some promising work on stochastic systems (\cite{lillacci2013signal}, \cite{sisson2007sequential}), there is only very little literature available on stochastic systems using trajectory data, and even the available methods are computationally very demanding (see for instance \cite{golightly2014bayesian}, \cite{golightly2011bayesian}, \cite{andreychenko2011parameter}, \cite{stathopoulos2013markov}). Further, these methods often rely on approximating the model dynamics  (for instance using the diffusion approximation (\cite{gillespie2000chemical}) or linear noise approximation (\cite{elf2003fast}) ), but these approximations may not always be justifiable (in the case of low copy numbers of the reactants for example) and might obscure crucial system behaviour.}

%{\it \color{red}High dimensional parameter space (Problem)}  
\par{One particular problem that is common to most inference methods is the usually high dimensional parameter space. Most of the sampling based inference techniques require the exploration of the full parameter space in order to find areas of ``good'' parameters, which is no easy task as the dimension of the parameter space increases.}

%{\it \color{red}Sum up of what we do}  
\par{We are proposing an inference method that relies on two main ideas. We utilize Dirichlet Process Gaussian Mixture Models for the estimation of the distribution of parameters (see for instance \cite{gorur2010dirichlet}). This technique is a sophisticated machine learning tool that shows very good performance even for high-dimensional spaces and is further very robust to outliers. We then employ a form of nested sampling, that iteratively samples parameter vectors from a sequence of distributions, that can be used to approximate the full posterior. The parameter vectors are sampled in an independent manner from the estimated densities making our method fully parallelizable. 
}

\par{In the following we introduce the model class we will be dealing with and give a brief overview over the task of parameter inference. Then we illustrate the mechanisms of our method and demonstrate its performance on high-dimensional deterministic as well as stochastic problems.} 

\subsection{Chemical Reaction Networks}
% {\it Markov Model}\\
\par{We are considering a $U$-dimensional Markov Process $X(t)$ depending on a $V$-dimensional parameter vector $\theta$. We denote with $X_{u}(\tau)$ the $u^{\textnormal{\tiny th}}$ entry of the state vector at time $\tau$ and with $X(\tau) = \{X_{u}(\tau)\}_{u = 1, \ldots, U}$ the state vector at time $\tau$. 
 
In the context of stochastic chemical reaction networks this Markov process describes the development of $U$ species $\bm{X}_1, \bm{X}_2, \ldots, \bm{X}_U$ through $R$ reactions $\mathcal{R}_1, \mathcal{R}_2, \ldots, \mathcal{R}_R$ written as 
$$\mathcal{R}_r = \sum\limits_{u}^U p_{ru}\bm{X}_u \rightarrow \sum\limits_{u}^Uq_{ru} \bm{X}_u,$$
where $p_{ru}$ and $q_{ru}$ are the numbers of consumed and produced species for reaction $r$.
The states $X_{u}(\tau)$ correspond to the number of molecules of species $\bm{X}_u$ at time $\tau$ and each reaction $\mathcal{R}_r$ has an associated propensity. The propensities as well as other properties of the model depend on a $V$-dimensional parameter vector $\theta$.

\par{ The other modelling approach we consider is the deterministic, where the state at time $\tau$ follows a deterministic dynamic, described by an ordinary differential equation (ODE)
$$\frac{\mathrm{d}}{\mathrm{d\tau}}{X}(\tau) = f(X(\tau), \tau, \theta), $$
with a function $f$ determined by the chemical reaction network. Such a formulation is particularly useful, when instead of single cell trajectories, one uses the average of many such trajectories, since the mean behaviour of a stochastic chemical reaction network can be described with an ODE. }

\subsection{General Task}
\par{Regardless of the chosen modelling, the process ${X}(t)$ is assumed to be not directly observable but only through a $P$-dimensional observation vector 
$$Y_\tau \sim p(\cdot | X_\tau, \theta).$$
The observations $Y_\tau$ at time $\tau$ depend on the state $X_\tau$ and on the $V$-dimensional parameter vector $\theta$. 

It is assumed that the variable $Y$ is not observed at all times but only on $T$ timepoints $\tau_1, \ldots, \tau_T$ and only for $M$ different trajectories. The observed state for trajectory $m$ and time point $\tau_t$ will be denoted with $y^m_t$. The $m^\textnormal{\tiny th}$ trajectory until the $t^\textnormal{\tiny th}$ time point will be denoted with $\bm{y}^m_{t} = \{y^m_1, \ldots, y^m_t\}$ and with $\bm{y} = \{\bm{y}_T^m\}_{ m = 1, \ldots, M}$ we denote all trajectories at all time points. If the observation is obtained by simulation using a certain parameter $\theta$ we will write $\bm{y}_\theta$. 

In the Bayesian approach the parameter vector $\theta$ is treated as a random variable with associated prior $p_0(\theta)$. The goal is not to find just one set of parameters, but much rather to compute the posterior distribution $p(\theta | \bm{y})$ of $\theta$
$$p(\theta | \bm{y}) \propto p(\bm{y} | \theta) p_0(\theta),$$
where $p(\bm{y}| \theta)$ is the likelihood of $\theta$ for the particular observation $\bm{y}$.
This has several advantages over a single point estimate as it gives insight over the areas of ``good'' parameters as well as about their relevance for the simulation outcome (a wide posterior indicates non identifiability for example). The challenge often lies in the fact that for stochastic systems the likelihood $p(\bm{y}|\theta)$ is not easily available and needs to be approximated. 

In this paper we follow the Bayesian approach and try to recover the posterior of the particle $p(\theta | \bm{y})$.
}

\section{Results}
In the following we first outline the proposed algorithm to give a general idea of our approach, before we describe it in full detail. \\
\par{ \bf Algorithm Outline}

\begin{algorithm}
\caption{Algorithm Outline. An overview over the involved steps of our proposed algorithm. The plue part illustrates the part that is parallelized.}
\begin{algorithmic}[1]
\STATE{Given observations $\bm{y}$, a prior $p_0(\theta)$ for $\theta$, and a sequence of thresholds $\epsilon_k$}
\STATE{sample particles from $p_0$}
	\FOR{k=1, \ldots }
		\STATE {Approximate super level set of particles $\theta$ with $p(\bm{y} | \theta) \geq \epsilon_k$ using {\color{green} previous particles},{ DP-GMM and rejection sampling}.}
		\STATE {\color{red} Independently sample new particles from approximated super level set.}
	\ENDFOR
\end{algorithmic}
\label{algo::LNS_overview}
\end{algorithm}

\begin{figure*}
\includegraphics[width=\textwidth]{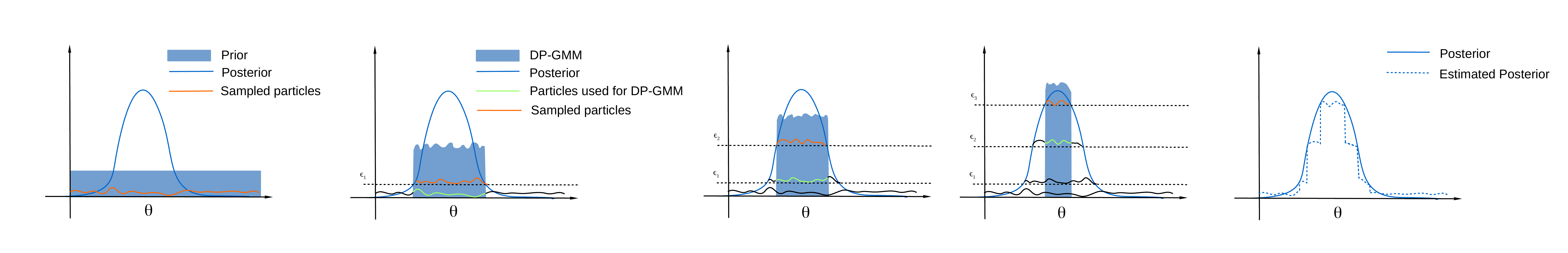}
\caption{Illustration of the algorithm. First particles (orange) are sampled from the prior (or previous DP-GMM estimation). Then these particles (those corresponding to a super level set of the likelihood, green) are being used to approximate the distribution of the super level set. Then the DP-GMM approximation of the super level set is used to sample a new set of particles corresponding to the super level set (orange). The sampling of the particles from the DP-GMM can be parallelized.}
\label{fig:algo_illustration}
\end{figure*}

\par{In this work we present a novel algorithm that reliably infers model parameters for large stochastic or deterministic models using trajectory data. We found a way to efficiently sample parameter vectors (particles) from the super level set of the likelihood (sets of particles with a likelihood equal to or higher than some threshold) corresponding to an increasing sequence of thresholds $\epsilon_k$. From these samples from different super level sets we can then recover the full posterior (see Supplementary Material II). A brief outline of the algorithm can be seen in Algorithm \ref{algo::LNS_overview} and a visualization fo the algorithm is shown in Figure \ref{fig:algo_illustration}.

We approximate these high dimensional super level sets through Dirichlet Process Gaussiam Mixture Models (DP-GMM) and rejection sampling (see Line 4 in Algorithm \ref{algo::LNS_overview}, see section \ref{sec:methods}). DP-GMM prove to be very efficient even with comparably few samples and are robust to outliers, which allows us to obtain a good approximation of the super level set, even when the likelihood of the underlying particles is not exactly known. The samples drawn for each super level set are independent (Line 5 in Algorithm \ref{algo::LNS_overview}), which allows us to parallelize the sampling of the particles. The sequential increase of the likelihood threshold allows us to efficiently explore the parameter space.}

\par{This exploration of the parameter space through super level sets of the likelihood is very similar to what is used in Nested Sampling approaches (as presented for instance in \cite{aitken2013nested} or  \cite{pullen2014bayesian}), which we will briefly outline. Nested Sampling relies on iteratively sampling particles and computing their corresponding likelihoods. At first the samples are drawn from the prior. From the initially sampled particles the particle with the lowest likelihood is rejected and a new particle is sampled and accepted only if its likelihood is higher than the current lowest likelihood. This approach results in a sequence of samples from different super level sets, from which the posterior can be recovered. The usual formulation of Nested Sampling replaces in each iteration the particle with the lowest likelihood with a new particle, which makes it hard to parallelize this approach. Further, the sampling of the new particle is challenging and usually involves some volume estimation or random walk and the knowledge of the exact likelihood (see for instance \cite{johnson2015sysbions},\cite{mukherjee2006nested} or \cite{feroz2013importance}) }

\par{Since for stochastic models, the exact likelihood is usually not available, we rely on a particle filter approximation of the likelihood (see for details Supplementary Materials I). Such likelihood approximations are frequently used in Monte Carlo Markov Chain (MCMC) methods (see e.g. \cite{golightly2011bayesian} or \cite{boys2008bayesian}). These methods construct samples from a Markov chain, whose stationary distribution is the posterior distribution of the parameters that we are interested in.
MCMC methods are exact in the sense that they generate samples from the exact posterior (up to the inaccuracy of the likelihood approximation), but suffer from several known weaknesses, such as a possibly long burn in time, possibly high autocorrelation and thus poor exploration of the parameter space, no straight forward parallelization and required extensive tuning by the user. } 

\par{Our proposed method is very well parallelizable and explores even high dimensional parameter spaces efficiently. In the following we elaborate on the details of the algorithm.} \\

\begin{figure*}
\includegraphics[width=\textwidth]{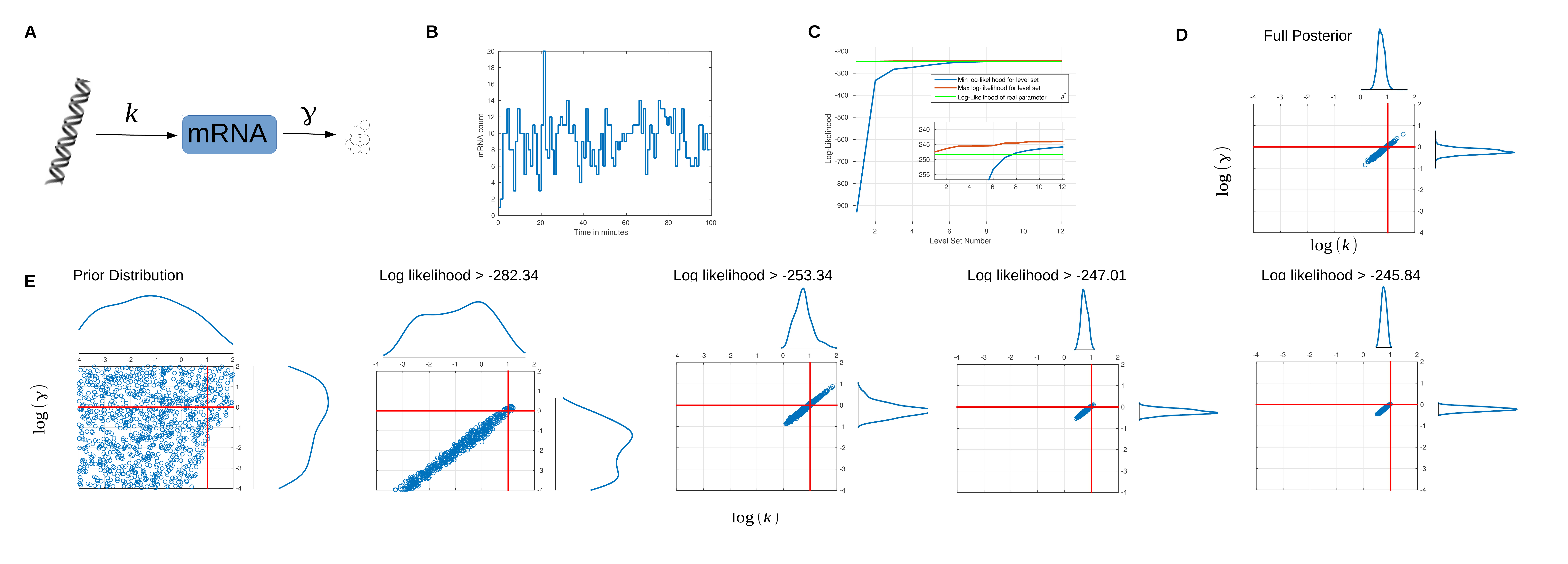}
\caption{{\bf A}: Schematic of the birth-death model. mRNA gets transcribed from DNA and degrades with fixed rate constants. {\bf B}: A single stochastically simulated trajectory of the birth-death system with birth rate $k = 10$ and degradation rate $\gamma = 1$. {\bf C:} The development of the smallest (blue) and largest (red) log-likelihood of the level-set for each iteration. The green line indicates the log-likelihood of the true parameter $\theta^*$ that was used for the simulation.. {\bf D}: The posterior of the log of the transcription and degradation rates as obtained with the LNS algorithm using only one simulated trajectory. The red lines indicate the true log-parameter values. {\bf E}:  The level sets of the log-likelihood of the log-translation and log-degradation rates as obtained with the LNS algorithm using only one observed trajectory.}
\label{fig:birth_death}
\end{figure*}

\par{ \bf Algorithm Details} \\
From now on we use the convention that for any variable $p$, $\hat{p}$ denotes its approximation. If $p$ is a distribution this approximation is usually a Dirichlet Process Gaussian Mixture approximation (see section \ref{sec:dpgmm}). This approximation corresponds to an approximation of the full distribution, while when $l$ is the likelihood, the approximation $\hat{l}$ is usually obtained using a particle filter approach (see section Supplementary material I) and corresponds to an approximation of the likelihood evaluated at a single point. \\

\begin{algorithm}
\caption{Likelihood Nested Sampling Algorithm}
\begin{algorithmic}[1]
\STATE{Given observations $\bm{y}$, a prior $p_0(\theta)$ for $\theta$, and some $0 < \alpha < 1$.}
\STATE{Set $\epsilon_1 = 0$, sample $N$ particles from $p_0$ to obtain the set $\mathcal{S}_0$}
	\FOR{k=1, \ldots }
		\STATE Estimate density $\hat{q}_{k-1}$ from the set $\mathcal{S}_{k-1}$
		\STATE Set $\mathcal{S}'_k = \mathcal{S}_{k-1}$
		\STATE Remove all particles $\theta$ from $\mathcal{S}'_k$ with $\hat{l}(\theta) < \epsilon_k$
		\STATE Estimate density $\hat{q}'_k$ from the set $\mathcal{S}'_{k}$
		\WHILE{number of particles in $\mathcal{S}'_{k}$ is less than N}
			\STATE sample $\theta \sim q'_{k} q_{k-1}^{-1} \approx L_{k}$. 
			\IF {$\hat{l}(\theta) \geq \epsilon_{k}$} 
				\STATE accept $(\theta_k^n)$ to $\mathcal{S}'_k$
			\ENDIF 
		\ENDWHILE
		\STATE Set $\mathcal{S}_k = \mathcal{S}'_k$
	\ENDFOR
\end{algorithmic}
\label{algo::LNS}
\end{algorithm}

The goal of the algorithm is to iteratively sample particles from the level set distributions
$$L_k(\theta) \propto \begin{cases}1 & \textnormal{if } l(\theta) > \epsilon_k \\ 0 & \textnormal{otherwise} \end{cases}$$
for different $\epsilon_k$. Since it is in general not possible to sample from the level sets directly, we create a sequence of sets $\mathcal{S}'_k$ which we use, combined with DP-GMM density approximation and rejection sampling, to sample from $L_k$. Once we have these level sets, we can recover the full posterior as described in the Supplementary material II.
We first sample a set $\mathcal{S}_0$ of $N$ particles 
$$\mathcal{S}_0 = \{\theta_i^0\}_{i = 1, \ldots, N}$$
 from the prior $p_0(\theta)$ and approximate for each $\theta_i$ the likelihood $l(\theta_i)$. The particles in set $\mathcal{S}_0$ are distributed according to the prior which we will denote with
$$\mathcal{S}_0 \sim p_0.$$
Our goal is to use this set to sample from the level set $L_1$. To this end we discard all particles with a likelihood lower than the threshold $\epsilon_1$ (which we choose to be the $\alpha_1$-quantile of the likelihoods in the set $\mathcal{S}_0$ for some $\alpha_1$) to obtain a set of $N'_1 = \lfloor(1-\alpha_1)N\rfloor$ particles 
$$\mathcal{S}'_1 = \{\theta_i^1 | \hat{l}(\theta_i^1) > \epsilon_1\}_{i = 1, \ldots, N_1'}.$$
The particles in this set are distributed according to $q'_1 :=p_0(\theta) L_1(\theta)$
$$\mathcal{S}'_1 \sim \underbrace{p_0(\theta)L_1(\theta)}_{q'_1}.$$
This set can be used to approximate the distribution
$$q'_1 = p_0(\theta)L_1(\theta) \propto \begin{cases}p_0(\theta) & \textnormal{if $\hat{l}(\theta) > \epsilon_1$} \\ 0 & \textnormal{otherwise} \end{cases}$$
by using DP-GMM (see \ref{sec:dpgmm}).
Now we can use rejection sampling to sample $\alpha_1N$ particles from $L_1(\theta)$, by sampling $\theta^*$ from $q'_1$ and accepting it with a probability proportional to $p_0(\theta^*)^{-1}$. 
Together with the original $N'$ particles we obtain the set
$$\mathcal{S}_1 \sim  \underbrace{\frac{N'_1}{N} q'_{1} + \frac{N-N'_1}{N} L_1}_{q_1}$$
of $N$ particles with a likelihood larger than $\epsilon_1$. 
Observe that this way we only have to sample $N - N'_1 = \lceil\alpha_1 N\rceil$ particles in each iteration which decreases the runtime drastically.
We repeat this procedure iteratively. {Assuming we have the set 
$$\mathcal{S}_k \sim q_k,$$
we sample $\lceil\alpha_{k+1}N\rceil$ particles from the level set $L_{k+1}$ by discarding all particles from set $\mathcal{S}_k$ with a likelihood lower than $\epsilon_{k+1}$ to obtain 
$$\mathcal{S}'_{k + 1} \sim \underbrace{\hat{q}_k L_{k+1}}_{q'_{k+1}}, $$
use DP-GMM to obtain an approximation  $\hat{q}'_{k+1}$ of $q'_{k+1}$ and generate samples from the level set $L_{k+1}$ by sampling $\theta^*$ from $\hat{q}'_{k+1}$ and accepting it with probability proportional to $\hat{q}_k(\theta^*)^{-1}$ ($\hat{q}_k$ is the DP-GMM approximation of $q_{k}$). The use of rejection sampling is crucial, since $\hat{q}_k$ as well as $\hat{q}'_{k+1}$ are DP-GMM approximations of sets of particles with only approximated likelihoods. Thus the rejection sampling accounts not only for a possible asymmetric prior, but also for any estimation error from the density or likelihood approximation (see Supplementary Material III).}

Consequently we also have
$$\mathcal{S}_{k+1} \sim \frac{N'_{k+1}}{N} \hat{q}'_{k+1} + \frac{N-N'_{k+1}}{N} L_{k+1}(\theta) =: q_{k+1}.$$ 
 In the following we demonstrate its performance on three chosen examples.\\

\par{ \bf The stochastic birth-death Model}
\par{We first demonstrate the basic mechanisms of the proposed method on the very simple birth-death model, illustrated in Figure \ref{fig:birth_death} A. This example serves well as a first illustration since it only has two parameters. In this model a single species (mRNA) is constitutively generated at a birth rate $k$ and degrades at a degradation rate $\gamma$. We simulate a single trajectory (Figure \ref{fig:birth_death} B) using the Stochastic Simulation Algorithm (\cite{gillespie1977exact}) taking $k = 10$ and $\gamma = 1$ and an initial mRNA molecule number of $0$. Using this simulated trajectory we infer the used parameters $k$ and $\gamma$. We assume the initial mRNA number to be known.  We assume a normal measurement noise with zero mean and a standard deviation of 2. At first our algorithm generates a set of parameter vectors (particles) that are sampled from a log-uniform distribution on the interval $[0.00001 ~100]\times [0.00001 ~100]$ and approximates the likelihood of each of those particles using a particle filtering approach (see Supplementary material I). If the likelihood could not be approximated (as for the particles in the lower right corner of the first graph of Figure \ref{fig:birth_death} E), the particles are not accepted. This can happen if the sampled particles result in simulations so far from the observed data, that the likelihood is approximated with 0. 

It then discards all particles with a likelihood smaller than a certain threshold which results in a discrete approximation of the super level set of the likelihood corresponding to that very threshold. By sampling particles from the current super level set and discarding all particles with a likelihood smaller than the next threshold, our algorithm generates a sequence of super level set distributions for the parameter (see Figure \ref{fig:birth_death} E). If we are interested in only one parameter vector that results in simulations close to our observation, we can pick one of the particles from the highest level set, otherwise we can also recover the full posterior as illustrated in Figure \ref{fig:birth_death} D. The resulting posterior is centred at the true parameter values, but its width indicates, that the parameters could not be identified uniquely, which is expected, since a wide range of parameters can result in the same trajectory. Figure \ref{fig:birth_death} C shows the development of the log-likelihood threshold and the highest achieved log-likelihood for each level set. As can be seen, the final thresholds are even higher than the log-likelihood of the true parameter. This is also not surprising, since the used trajectory is only one realization of the stochastic systems and certain other parameters may be just as likely to have resulted in the same trajectory.
We would like to point out that the super level sets illustrated in Figure \ref{fig:birth_death} E seem to have an ellipsoid shape, as we would expect in this case.} \\

\begin{figure*}
\includegraphics[width=\textwidth]{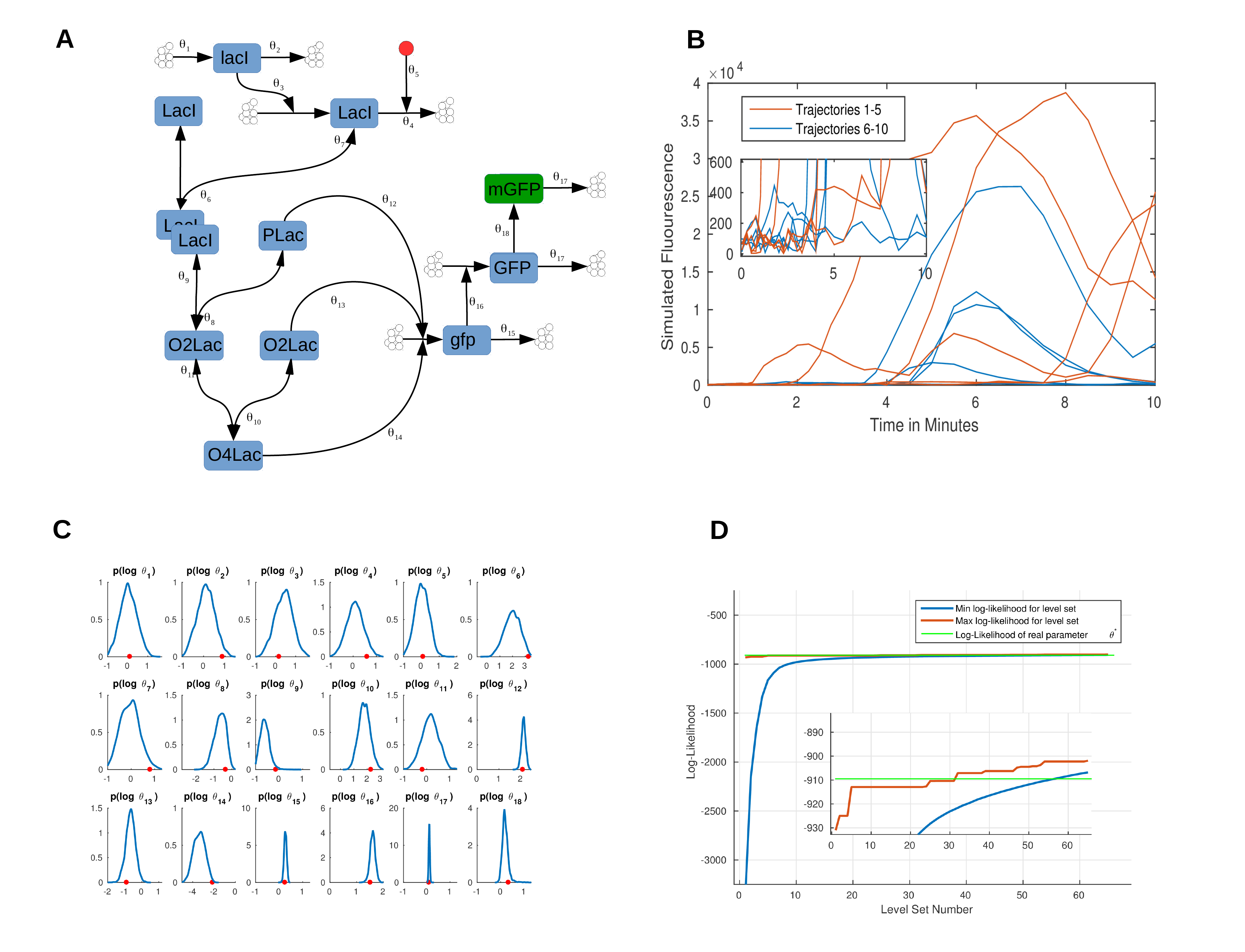}
\caption{{\bf A}: The Lac-Gfp model, containing 9 species and 18 reactions. {\bf B}: 10 Simulated model trajectories that were used for parameter inference. The system exhibits stochastic switch behaviour. {\bf C}: Inferred posterior of the logarithm of the parameters. The red dots indicate the true parameter value $\theta^*$ that was used for the simulation of the data. As prior, a log-normal distribution in the indicated bounds was used. {\bf D}: The development of the smallest (blue) and largest (red) log-likelihood of the level-set for each iteration. The green line indicates the log-likelihood of the true parameter $\theta^*$ that was used for the simulation.}
\label{fig:full_lac}
\end{figure*}

\par{\bf A large stochastic model: The Lac-Gfp model}
\par{We demonstrate how our algorithm deals with a realistic sized stochastic model, by inferring the posterior for the parameters of the LacGfp model illustrated in Figure \ref{fig:full_lac} A. This model has been already used in \cite{lillacci2013signal} as a benchmark, although with distribution-data (for details see Supplementary material and \cite{lillacci2013signal}). Here we use the model to simulate a number of trajectories and illustrate how our approach infers the posteriors of the used parameters. This model is particularly challenging in two aspects. First, the number of parameters is 18, making it a fairly large model to infer. Secondly, the model exhibits switch-like behaviour which makes it very hard to approximate the likelihood of such a switching trajectory (see the Supplementary material for a further discussion). In Figure \ref{fig:full_lac} we show 10 of the simulated trajectories. As can be seen, most trajectories exhibit noise fluctuations first and at some point ``switch on''. The switching occurs randomly and the observed fluorescence increases rapidly in orders of magnitude. This raises considerable problems for the likelihood estimation. The measured species in this example is fluorescent GFP where it is assumed that each GFP-molecule emits fluorescence according to a normal distribution.  We used 5 trajectories to infer the posterior, whose marginals are shown in Figure \ref{fig:full_lac} C. As can be seen, the posterior seems to capture the real parameter fairly well. It is also seen that some parameters seem to have a wider marginal posterior than others. This indicates that the model is more sensitive to those parameters with narrower marginal posterior, since the likelihood seems to be more sensitive in respect of those parameters. The development of the log-likelihood threshold illustrated in Figure \ref{fig:full_lac} D shows similar behaviour as for the birth-death model. The threshold eventually surpasses the actual log-likelihood of the real parameter $\theta^*$. The first time that a particle is found with a higher log-likelihood than $\theta^*$ is after 30 iterations. }
\\

\begin{figure*}
\includegraphics[width=\textwidth]{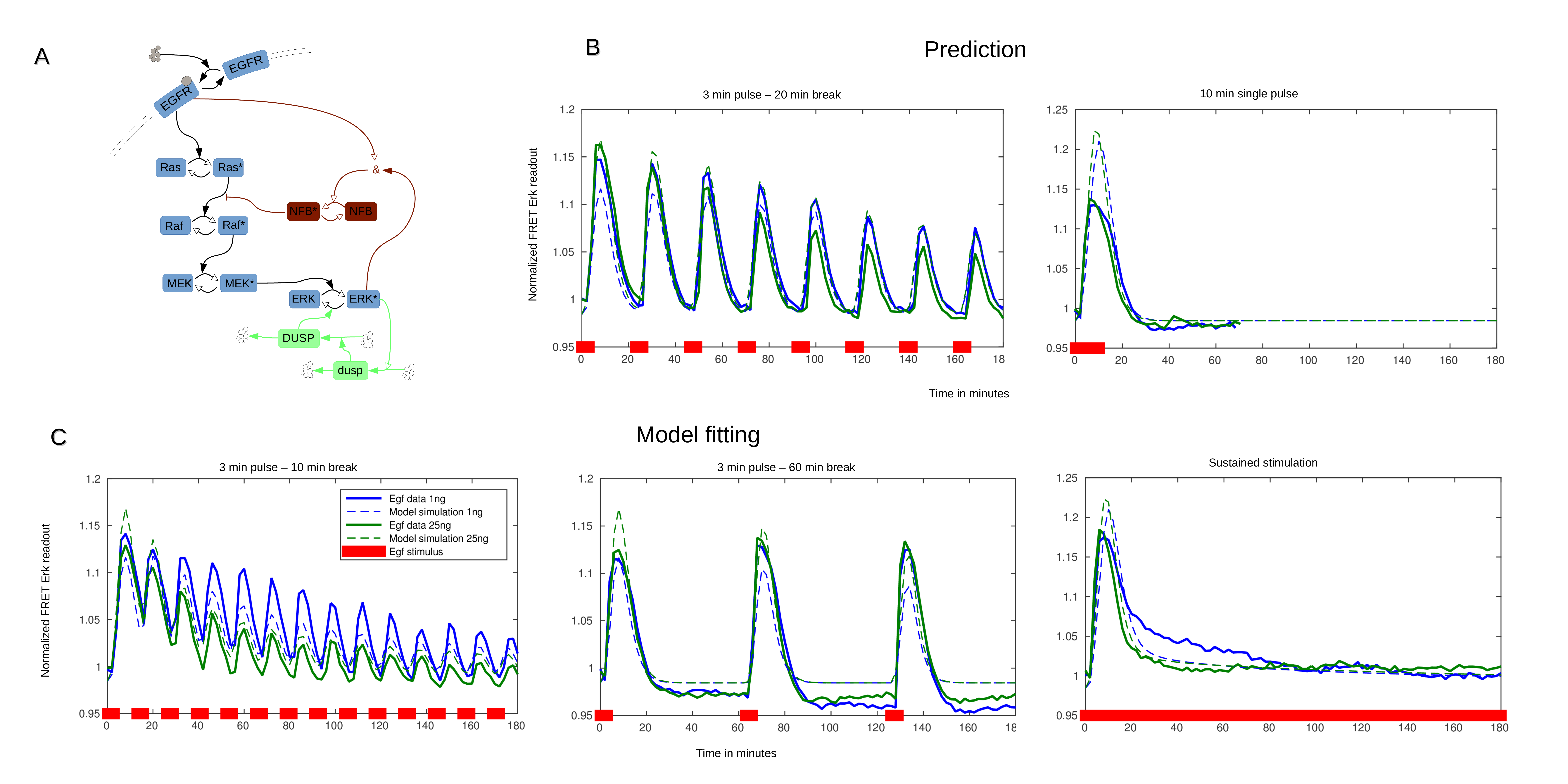}
\caption{  {\bf A}: Schematic representation of the Ras-Raf-Mek-Erk pathway model from \cite{ryu2015frequency}. Upon stimulation with Egf, the Egf-Receptor activates Ras, which in turn activates Raf, Mek and Erk. Dusp transcription is stimulated by phosphorilated Erk, while at the same time Dusp dephosphorilates Erk, making it a slow negative feedback from Erk upon itself. The NFB species is an unspecified, Erk and Egf dependent  negative feedback species. The model has a total of 36 unknown parameters and 11 species. {\bf B}: Prediction of the model. The particle with the highest likelihood was used to simulate an input stimulus, previously assumed unknown for the model inference. {\bf C}: Fitting of the Egf-Ras-Raf-Mek-Erk model. The inference was performed using these 6 trajectories. The particle with the highest likelihood was then used for the simulation.}
\label{fig:full_egf}
\end{figure*}

\par{\bf A realistic example: The deterministic Raf-Mek-Erk signalling cascade}
\par{Our method can similarly be applied to deterministic models. In a deterministic setting, the hidden system states $X_\tau$ do not develop according to a Markov Chain but according to a set of ordinary differential equations that depend on the parameter vector $\theta$. Again, it is assumed that the states are not directly observable but only through noisy measurements $Y_\tau \sim p(\cdot | X_\tau, \theta)$. In the deterministic case the computation of the likelihood simplifies, since we do not need a particle filter but instead can solve the ordinary differential equations of the model with the particular parameters and compute the likelihood from the resulting trajectory. The use of DP-GMM density estimation allows us to deal with high dimensional parameter spaces and the sequential nature of our algorithm makes a parallelization straight forward. We demonstrate our method on the example of the Ras-Raf-Mek-Erk pathway (Figure \ref{fig:full_egf}). We use the model introduced in \cite{ryu2015frequency}, which consists of an activation cascade formed by the Egf receptor, Ras, Raf, Mek and Erk (see Figure \ref{fig:full_egf} C). The model also features a slow negative feedback through Dusp and a fast receptor dependent negative feedback. The read out is the amount of phosphorylated Erk as measured by a FRET sensor. For details on the FRET sensor see \cite{ryu2015frequency}. As training data for our parameter inference we used the observed mean behaviour of the phosphorylated Erk upon different stimulations, see Figure \ref{fig:full_egf} A. In order to check our inferred posterior, we picked the particle with the highest likelihood from the posterior and simulated the model for different stimuli. Figure \ref{fig:full_egf} A shows the model behaviour for stimuli that have been used to identify the system, while Figure \ref{fig:full_egf} B shows the model simulations for stimuli that have not been used for the inference. We see that using the inferred particle the model is able to reproduce the training data as well as predict the systems behaviour to new stimuli. We can use the inferred posterior to draw biological conclusions that, as a future step, could be verified experimentally. For instance, one of the model assumption is that dusp gets transcribed constitutively at a certain rate $\textnormal{dusp}_{\textnormal{basal}}$ and induced by Erk with the rate $\textnormal{dusp}_{\textnormal{ind}}$. Along all other parameters, both rates were inferred from the provided data as well as the dusp and DUSP initial concentrations. Their marginal posterior distribution is shown in Figure \ref{fig:dusp_post}. From a quick look at the marginal posteriors we see that in order for the model to produce the observed behaviour, the constitutive dusp transcription needs to be four orders of magnitude smaller than the Erk induced transcription. We further see that the initial concentration of dusp and DUSP needs to be very low ($\sim  10^{-3}$) compared with the initial concentration of Erk (0.6) for example. We mention these particular observations to illustrate that given the posterior, we can deduce biological statements that can be experimentally checked in order to investigate the validity of the model.}

\begin{figure}
\includegraphics[width=0.5\textwidth]{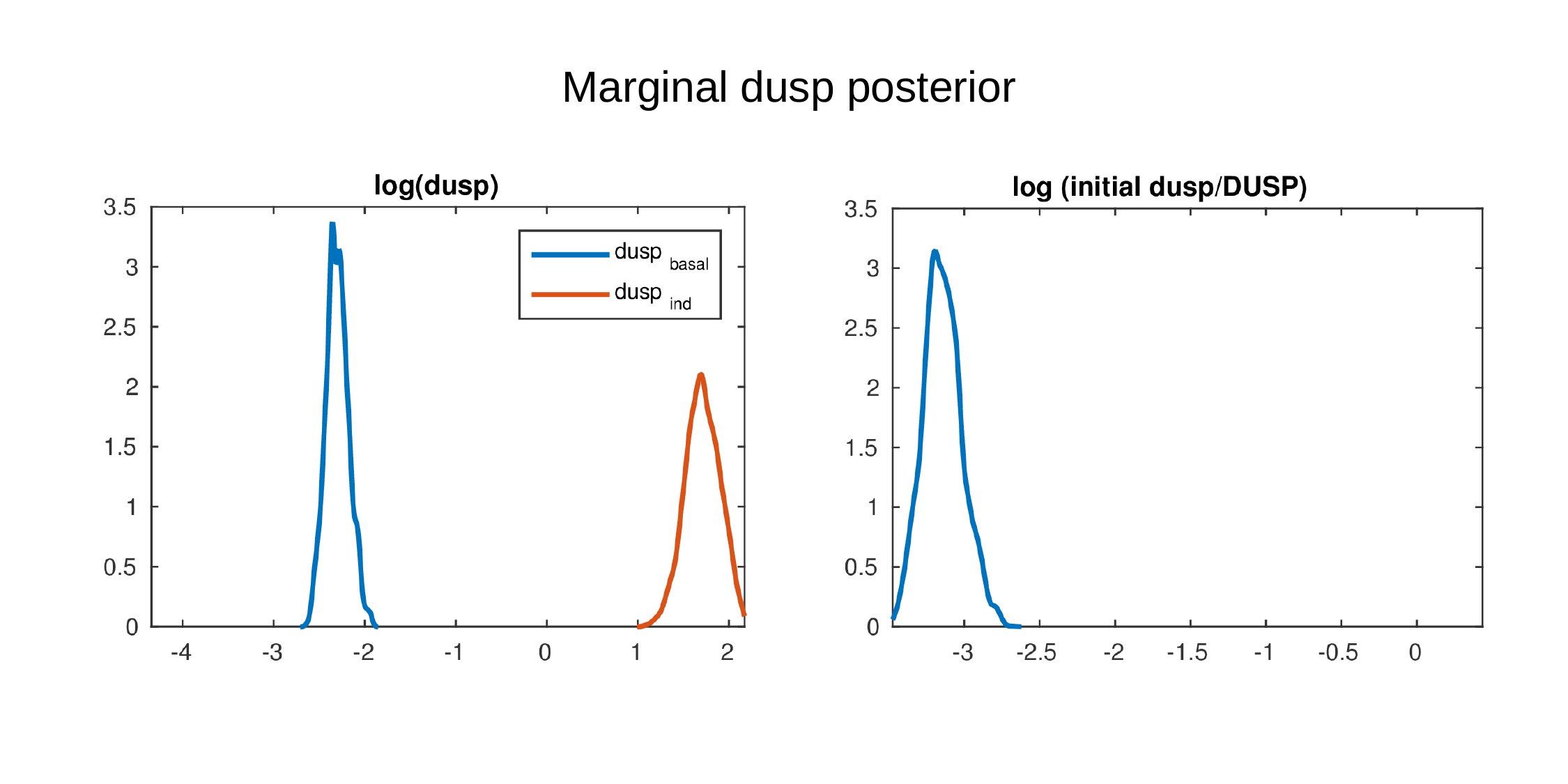}
\caption{The marginal posterior distribution for the Egf-Ras-Raf-Mek-Erk model of the constitutive transcription rate $\textnormal{dustp}_{\textnormal{basal}}$, the Erk induced dusp transcription $\textnormal{dusp}_{\textnormal{ind}}$ and the initial concentration of dusp and DUSP (which is assumed to be equal).}
\label{fig:dusp_post}
\end{figure}

\section{Conclusion}
\par{ We have introduced an inference method that reliably infers the posterior for model parameters of large stochastic and deterministic systems. We found a way to sample directly from the super level set of the likelihood without needing the exact likelihood and having to resort to volume approximations or random walks. This also allows us to parallelize the sampling of the super level sets. 
We have demonstrated the performance of our method applied to large stochastic models as well as realistic sized deterministic models and have shown how the obtained posterior of particles can be used not only to simulate and predict the model behaviour, but also to create biological statements that can be checked in order to verify model consistency.}

\section{Remarks}\label{sec:methods}

In the following we briefly outline some particular relevant details of the implementation of the algorithm. \\

{\it Use quantiles for the computation of $\epsilon_{k+1}$} 

To determine the sequence $\{\epsilon_{k}\}_k$, we use the $\alpha$-quantile of $q_{k}$ for some $0< \alpha < 1$, where we choose the same $\alpha$ for all $k$,
 $$\alpha_k = \alpha, \quad \forall k.$$
 Thus from the set $\{\theta_i^k\}_{i= 1, \ldots, N}$ only $\lceil\alpha N\rceil$ particles will have a likelihood that is lower than $\epsilon_{k+1}$ and only these particles will need to be resampled. \\

{\it DP-GMM for density estimation }\label{sec:dpgmm}

In each iteration the density $q'_{k}$ has to be approximated by $\hat{q}'_k$ from a set of $N$ particles $\{\theta_i^k\}_{i = 1, \ldots, N}$. This is a task that needs to be performed for most Monte Carlo inference algorithms and is usually done with kernel density estimation (KDE) which places a perturbation kernel on each particle and takes $\hat{q}'_k$ to be the sum of all these kernels.  This approach has several drawbacks as it is not clear what kernel to use or how to choose its width. Further, KDE is sensitive to outliers and shows a poor performance with sparse data. This poses a serious problem for most sample based inference techniques, since in most applications the parameter space is large and only comparably few particles are available.

To overcome these problems we propose to use a density estimation technique that approximates the target distribution with a mixture of Gaussians, where the number, shape and weight of each Gaussian is inferred from the particle population. The estimation technique - Dirichlet Process Gaussian Mixture Model (DP-GMM) (\cite{gorur2010dirichlet}) - uses a hierarchical prior on the mixture model and assumes that the mixture components are distributed according to a Dirichlet Process. The inference of the distribution is an iterative process that uses Gibbs sampling to infer the parameters and hyper parameters of the mixture model. This procedure is usually computationally demanding. However, since only $\lceil\alpha N\rceil$ particles are resampled in each population, the previous parameters and hyper parameters provide an already good prior estimate for the new estimation which allows to find a new estimation with only few iterations. DP-GMM estimations perform comparably well with sparse and high dimensional data, because they use the joint information of all samples to infer the high dimensional density. The use of DP-GMM is what enables us to apply our algorithm even to high dimensional parameter spaces. It also allows us to follow a nested sampling approach without needing the exact likelihood. Even when we assume that we do have the exact likelihood, which would reduce our algorithm to the already established nested sampling method, the use of the DP-GMM estimation allows us to efficiently sample from a high dimensional level set, without having to approximate the level set directly.

Further details are given in the Supplementary material and can also be found in \cite{gorur2010dirichlet}, \cite{rasmussen1999infinite} and \cite{teh2006hierarchical} (in particular \cite{gorur2010dirichlet} shows a comparison between DP-GMM and KDE). 

\bibliographystyle{plain}
\bibliography{bibliography}

\onecolumngrid
\part*{Supplementary}
\section{Computing the Likelihood}\label{sec:sup likelihood}

\subsection{Computation of the likelihood}\label{sec:likelihood}

The most critical part in the algorithm is the computation of the likelihood, which is why we will quickly discuss one of the main issues and our approach of solving it. We follow the particle filter approach from  \cite{golightly2011bayesian}, which we will briefly outline.

 The likelihood for the set of trajectories $\{\bm{y}^m\}_{m = 1\ldots, M}$ is computed as the product of the likelihoods of each trajectory
$$p(\bm{y} | \theta) = \prod_{m = 1}^M p(\bm{y}^m | \theta).$$
The likelihood for each trajectory $\bm{y}^m$ is computed as the product of the likelihoods at each timepoint
$$p(\bm{y}^m | \theta) = \prod_{t=1}^T p(y_t^m | \bm{y}_{t-1}^m \theta).$$
To be precise, the likelihood for each trajectory up until timepoint $t+1$, $\bm{y}_{t + 1}^m$ gets computed recursively
$$\hat{p}(\bm{y}_{t+1}^m | \theta) \approx \hat{p}(\bm{y}_t^m | \theta)\int p(y_{t+1}^m | X_{t+1})p(X_{t+1} | \bm{y}_t^m,\theta)d X_{t+1}.$$
The integral in the recursive formula is approximated by forward simulation of $H$ hidden trajectories which are afterwards resampled. In particular, let us assume we have an empirical distribution of $N$ states from the initial state distribution $X_0^i \sim p_{X_0}$, $i = 1, \ldots, N$. To obtain an approximation of the distribution $p(X_1 | \bm{y}_0)$, we weight each of the $H$ states $X_0^h$ with the likelihood $w_0^h = p(\bm{y}_0 | X_0^h)$ and resample them using those weights to obtain an empirical approximation for the distribution $p(X_0 | \bm{y}_0, \theta)$. By forward simulating from $X_0^h$, we thus get the states $X_1^h \sim p(X_1 | \bm{y}_0, \theta)$. The likelihood for $\bm{y}_1$ can then be approximated by
\begin{equation}\label{eq:likelihood approx} p(\bm{y}_1 | \theta) = p(\bm{y}_0|\theta) \frac{1}{H}\sum_h p(\bm{y}_1 | X_1^h).\end{equation}

This procedure of weighting, resampling and forward simulating allows us to approximate for each $t$ the distribution $p(X_{t+1} | \bm{y}_t^m, \theta)$. Once this is accomplished, the computation of $p(y_{t+1}^m | X_{t+1})$ is straight forward. This procedure results in Algorithm \ref{algo:particleFilter}.

\begin{algorithm}
\begin{algorithmic}
\STATE{Given $H$ states $X_0^i$ sampled from some initial distribution $p_{X_0}$ and a parameter vector $\theta$.}
\STATE{Compute for each $i$ the likelihood $l_0^i = p(\bm{y}_0 | X_0^h)$.}
\STATE{Set $p(\bm{y}_0 | \theta) = \frac{1}{H} \sum_{i=1}^H l_0^i$.}
\STATE{Compute weights $w_0^i = \frac{l_0^i}{\sum_{i = 1}^H l_0^i}$.}
	\FOR{t=1, \ldots, T }
		\STATE Sample $H$ indices $i'$ according to weights $w_{t-1}^i$.
		\STATE Simulate $X_{t}^i$ starting from $X_{t-1}^{i'}$ and using $\theta$.
		\STATE Compute $l_t^i = p(y_t | X_t^i, \theta)$.
		\STATE Compute $p(\bm{y}_t | \theta ) = p(\bm{y}_{t-1} | \theta)\frac{1}{H} \sum_{i=1}^H l_t^i $.
		\STATE Compute  weights $w_0^i = \frac{l_0^i}{\sum_{i = 1}^H l_0^i}$.
	\ENDFOR
\end{algorithmic}
\caption{Likelihood Approximation using particle filter}
\label{algo:particleFilter}
\end{algorithm}

\section{Approximating the full posterior}\label{sec:full posterior}
In Bayesian analysis the goal is to infer the full posterior $p(\theta | \bm{y})$ and not just the level sets $L_{\epsilon_k}(\theta)$. One obvious way of recovering the full posterior is to divide the parameter space into bins and take the average of all computed likelihoods in that bin. The number of available particles will be in general large, since our algorithm relies on sampling a large number of particles. This approach may, however, result in a very noisy approximation of the posterior, since the likelihoods are only approximated and gets quickly unwieldly as dimensions increase. We propose a different approach of recovering the likelihood by exploiting the approximation of the level sets $L_{\epsilon_k}(\theta)$. 

\subsection{Recovering the posterior from level sets}\label{sec:full posterior analytical}
At first we assume we have exact level set distributions for the thresholds $\epsilon_k$
$$L_k(\theta)\propto \mathbbm{1}_{p(\bm{y} | \theta) \geq \epsilon_k}.$$
Considering the normalization constant we have the equality
$$L_k(\theta) = \frac{\mathbbm{1}_{p(\bm{y} | \theta) > \epsilon_k}}{\int\limits_{\Omega} \mathbbm{1}_{p(\bm{y} | \theta) > \epsilon_k}d\theta}.$$
Writing $g(\epsilon_k) = \int\limits_{\Omega} \mathbbm{1}_{p(\bm{y} | \theta) > \epsilon_k}d\theta$,  we can obtain the likelihood from the level sets by
$$\int\limits_{0}^{\infty} L_k(\theta) g(\epsilon_k) d \epsilon_k  = \int\limits_{0}^{\infty} \mathbbm{1}_{p(\bm{y}|\theta) > \epsilon_k }(\epsilon_k) d\epsilon_k = \int\limits_{0}^{p(\bm{y} | \theta)} d\epsilon_k = p(\bm{y} | \theta).$$
We can multiply this likelihood by the prior $p(\theta)$ to obtain the posterior
$$p(\theta |\bm{y}) \propto p(\bm{y} | \theta) p(\theta).$$

\subsection{Use approximated level sets}
To obtain an approximation of the function $L_k(\theta)$,  we first need an approximation $\hat{p}(\bm{y}|\theta)$ for the likelihood $p(\bm{y}|\theta)$ in order to obtain a set of particles $\{\theta_i | \hat{p}(\bm{y}|\theta_i) > \epsilon_k)\}_{i = 1, \ldots, N}$.  This set of particles is then used to obtain an approximation $\hat{L}_k(\theta)$. In \ref{sec:sup likelihood} we explained how we obtain $\hat{p}(\bm{y}|\theta)$. 
The posterior can then be obtained through the approximation of the integral
$$\hat{p}(\theta | \bm{y}) =  \sum\limits_{k = 1}^K \hat{L}_k(\theta)p(\theta)\hat{g}(\epsilon_k)\delta_{\epsilon_k} \approx \int L_k(\theta)g(\epsilon_k)p(\theta)d \epsilon_k,$$
where
$$\delta_{\epsilon_k} = \begin{cases} \epsilon_k & \textnormal{if } k =1 \\ \epsilon_k - \epsilon_{k-1} & \textnormal{otherwise}\end{cases}.$$
We approximate $g(\epsilon_k)$ with $\hat{g}$ which we assume to be some appropriate approximation. We want $\hat{g}$ to be proportional to $\int\limits_{\Omega} \mathbbm{1}_{p(\bm{y} | \theta) > \epsilon_k}d\theta$. Since the volume of the confidence ellipsoid of the data is proportional to the product of the square roots of the eigenvalues of the covariance matrix of the data, we choose 
$$\hat{g} = \sqrt{ \det( \textnormal{Cov} L_k)}$$
where $\textnormal{Cov} L_k$ is the Covariance matrix of the distribution $L_k$. We like to note that in the case where the level sets are expected to be multi modal, this approximation may not be the best choice and one has to thing of different ways of approximating $g$, however in our encountered examples $ \sqrt{\det( \textnormal{Cov} L_k)}$ proved to be a sufficiently good approximation.

\subsection{Sampling from the posterior}
We described how, in the course of the algorithm run, we sample from approximations of each level set $\hat{L}_k$. Since we can sample from the level set $\hat{L}_k$, we can also sample from the weighted sum
$$ \sum\limits_{k = 1}^K \hat{L}_k(\theta)\hat{g}(\epsilon_k)\delta_{\epsilon_k} $$
and weighting each sample $\theta$ with $p(\theta)$ to obtain weighted samples form the posterior.

\subsection{Verification of approximated posterior}

To illustrate the validity of the recovered posterior, we consider a very simple RNA transcription model consisting of the species in table \ref{tab:rna_species}.

\begin{table}[tb]
\label{tab:rna_species}
\caption{Species and initial numbers for the RNA model}
\begin{tabular}{l l c }

\hline \hline

Species & Notation %& Unit 
& Initial Distribution\\

\hline

Gene in on state & $\textnormal{G}_{on}$ 
& fixed to 0 \\
Gene in off state & $\textnormal{G}_{off}$ 
&fixed to 1 \\
mRNA & RNA
&fixed to 0 \\
\hline

\hline
\end{tabular}

\end{table}

The reactions read as follows: \\

\textbf{Reactions}
\begin{enumerate}
\begin{small}
\item $ G_\textnormal{off} \stackrel{\theta_{1}}{\longrightarrow}G_\textnormal{on} $. Gene switch on.

\item $ G_\textnormal{on} \stackrel{\theta_{2}}{\longrightarrow}G_\textnormal{off} $. Gene switch off.

\item $G_\textnormal{on}  \stackrel{\theta_{3}}{\longrightarrow} G_\textnormal{on}  + RNA$. Transcription of mRNA.

\item $RNA\stackrel{\theta_4}{\longrightarrow} \emptyset $. mRNA degradation.
\end{small}
\end{enumerate}
The parameters with corresponding priors and values used for the simulation are shown in Table \ref{tab: priors rna}. 
\begin{table}[tb]
\centering

\caption{Prior distributions of the RNA-system parameters and the real values $\hat{\bm{\theta}}$ used for the simulation of $\bm{y}$.}
	
\begin{scriptsize}
\begin{tabular}{c l c c}

\hline \hline

Parameter & Meaning &  Prior interval on $\theta_i$& $\hat{\bm{\theta}}$ \\

\hline

$\theta_1$ & Rate of switching Gene on& [0.000001, 0.006] & 0.001\\

$\theta_2$ & Rate of switching Gene off&assumed known & 0.002\\

$\theta_3$ & mRNA transcription rate &assumed known  & 0.06\\

$\theta_4$ &mRNA degradation rate & assumed known & 0.002\\

\hline

\hline
\label{tab: priors rna}
\end{tabular}
\end{scriptsize}
\end{table}
We used this system to simulate one trajectory, observing the development of mRNA. The trajectory was observed on 25 equally spaced timepoints between 0 and 200 time units. We used the FSP \cite{munsky2006finite} to compute the (almost) actual posterior distribution of $\theta_1$. We used this (almost) exact likelihood to perform the parameter inference and show in Figure  \ref{fig:rna exact} the histogram of the recovered posterior, where the red dot indicates to position of the real parameter. As can be seen the posterior is nicely recovered from the level sets. 

%\begin{figure}
%\centering
%\includegraphics[width=0.5\textwidth]{somewhat_good_posterior_approx.eps}
%\caption{The result of a LNS run applied to the RNA System model, inferring the normalized posterior $\theta_1$ (histogram) and the actual posterior of $\theta_1$ as obtained with the FSP (red).}
%\label{fig:rna exact}. 
%\end{figure}

\section{Rejection Sampling}\label{sec:rejection}
In the process of the algorithm we need to sample particles from the level set $L_k$ using the set
$$\mathcal{S}'_k \sim \underbrace{q_{k-1} L_{k}}_{q'_k}$$
to construct the set $\mathcal{S}_{k}$. We have no means to sample directly from $L_k$, but we have the Dirichlet Process Gaussian Mixture Model density estimations $\hat{q}'_k$ and $\hat{q}_{k-1}$ of the densities $q'_k$ and $q_{k-1}$ respectively. 
We generate samples from $L_{k}$ by drawing a sample $\theta^*$ from $\hat{q}'_k$ and accept it with a probability $ \frac{M_k}{q_{k-1}(\theta^*)}$ with an appropriately chosen $M_k$. This guarantees that the sampled particle $\theta^*$ is distributed according to 
$$\theta^* \sim  \frac{\hat{q}'_{k} M_k}{\hat{q}_{k-1}} \approx M_k L_k$$
(with a slight abuse of notation since the right hand side is not normalized).  Since we only have approximations of $q'_{k}$ and $q_{k-1}$ and we cannot expect their ratio to be the actual level set, we also approximate the likelihood $l(\theta^*)$ and reject the particle if the approximated likelihood $\hat{l}(\theta^*)$ is smaller than the threshold $\epsilon_k$. The sampling of the particles is a crucial step in the LNS algorithm, which is why we will discuss the two important parts of our rejection sampling approach
\begin{itemize}
\item Choice of $M_k$
\item DP-GMM approximation
\end{itemize} 

\subsection{Choice of $M_k$}
This rejection sampling approach is guaranteed to give us the right target distribution $L_k$ whenever $M < q_{k-1}(\theta^*)$. Thus $M$ determines the domain from which we want to sample. If the particle $\theta^*$ (that is sampled from $\hat{q}'_k)$ is outside of this domain, we reject it automatically. 
For numerical reasons it is favourable to pick $M_k$ not too small. Particularly for $\theta$ where $q'_k(\theta) = q_{k-1}(\theta) = 0$ but $\hat{q}'_k(\theta) \neq  0 \neq \hat{q}_{k-1}(\theta)$, choosing $M_k$ too small may cause large numerical errors. In practice we did not observe a huge influence of the value for $M_k$ on the algorithm performance as long as it was smaller than $\max \limits_{\theta} \hat{q}_{k-1}$. For our algorithm runs we chose $M_k$ to be the value of the $5 \%$ quantile of $\hat{p}_{k-1}$. The main function of the rejection sampling is guaranteeing that estimation errors of $\hat{q}'_{k}$ and $\hat{q}_{k-1}$ don't propagate through iterations. For an illustration on the RNA model see Figure \ref{fig:rejection_example}. 

\subsection{DP-GMM Approximation}
The distributions $q'_k$ and $q_{k-1}$ are approximated using Dirichlet Process Gaussian Mixture Model approximation. This means that each of the distributions is approximated using Gaussian mixtures. This may at first glance not seem like an appropriate way of approximating level sets due to their non-smooth nature. However, DP-GMM estimations work very well in high dimensional spaces and since we use rejection sampling to obtain samples from the level set by sampling from the DP-GMM estimation, the estimation error does not get propagated through iterations. 

%\begin{figure}
%\centering
%\includegraphics[width=0.5\textwidth]{rejection_sampling_illustration.eps}
%\caption{The plots of the distributions $\hat{q}'_k$ (green), $\hat{q}_{k-1}$ (red) and the distribution obtained with rejection sampling (black). }
%\label{fig:rejection_example}. 
%\end{figure}

\section{Termination Criterion}\label{sec: termination criterion}
In the presentation of the algorithm we did not provide a termination criterion. An investigation of the development of the minimal and maximal log-likelihood threshold of inference runs for one trajectory and different values for $\alpha$ and different number of particles for the likelihood computation suggest two possible termination criteria. 
\begin{itemize}
\item {\bf The difference between minimal and maximal log-likelihood for a level set}\\
As the algorithm proceeds the difference between the maximal and the minimal log-likelihood for each level set decreases (see for instance Figure \ref{fig:lac_diff_post} B or ...). In general we cannot expect this difference to become 0 due to the fact that we always deal with an approximation of the likelihood. One way of defining a termination criterion is to define a threshold and as soon as the difference of the maximal and minimal log-likelihood is smaller than the predefined threshold the algorithm will stop. The difference of the maximal and minimal likelihood seems to be a monotonically decreasing function. Assuming that there exists a maximal log-likelihood on the parameter space, the difference is guaranteed to approach 0 (but maybe not monotonically). This way the algorithm is guaranteed to terminate. 
\item {\bf A low acceptance rate}\\
It is an interesting observation, that the acceptance rate for each iteration drops as the algorithm progresses. This brings us to the question how we would expect the acceptance rate to behave. At iteration $k$ the new particle $\theta^*$ is sampled from the approximative distribution
$$\theta^* \sim \hat{L}_k(\cdot) \approx L_k(\cdot) = \mathbbm{1}_{p(\bm{y} | \theta^*) > \epsilon_k}(\cdot)$$
and is accepted if $\hat{p}(\bm{y} | \theta^*) > \epsilon_k$. Thus we can expect that most particles will be accepted and that the acceptance rate depends on the accuracy of the approximations $\hat{L}_k$ and $\hat{p}(\bm{y} | \theta^*)$. In practise the approximation of $\hat{L}_k$ is done by DP-GMM and rejection sampling, whose accuracy does not depend on the iteration number. This leaves the accuracy of the likelihood approximation as the main contributor for the change in acceptance rate. More precise, the acceptance rate depends on the quotient of the distance of the likelihood of $\theta^*$ to the threshold $\epsilon_k$ and the variance of the likelihood estimation $\hat{p}(\bm{y} | \theta^*)$
$$\frac{p(\bm{y} | \theta^*) - \epsilon_k}{\textnormal{var}~\hat{p}(\bm{y} | \theta^*)}.$$
In Figure \ref{fig:lac_diff_acceptance} B, the acceptance rate for each log-likelihood threshold is plotted for different numbers of particles used for the likelihood approximation. As can be seen, the higher the number of particles (and thus the accuracy of the log-likelihood approximation) the higher is the acceptance rate.

Thus the acceptance rate depends on the variance of the likelihood estimation $\hat{p}(\bm{y}|\theta^*)$, which can be assumed to be independent of particular iteration number $k$ and on the average distance of the sampled particles $\theta$ to the current threshold $\epsilon_k$. If the acceptance rate drops drastically, it can therefore be assumed that all particles are very close to the current threshold, which is an indicator for the local optimality. An illustration of the dropping acceptance rate is seen in Figure \ref{fig:lac_diff_acceptance} where the acceptance rate is plotted against the threshold $\epsilon_k$ for different values of $\alpha$ ({\bf A}) and for different number of particles for the log-likelihood approximation ({\bf B}).

\begin{figure}
\centering
\includegraphics[width=\textwidth]{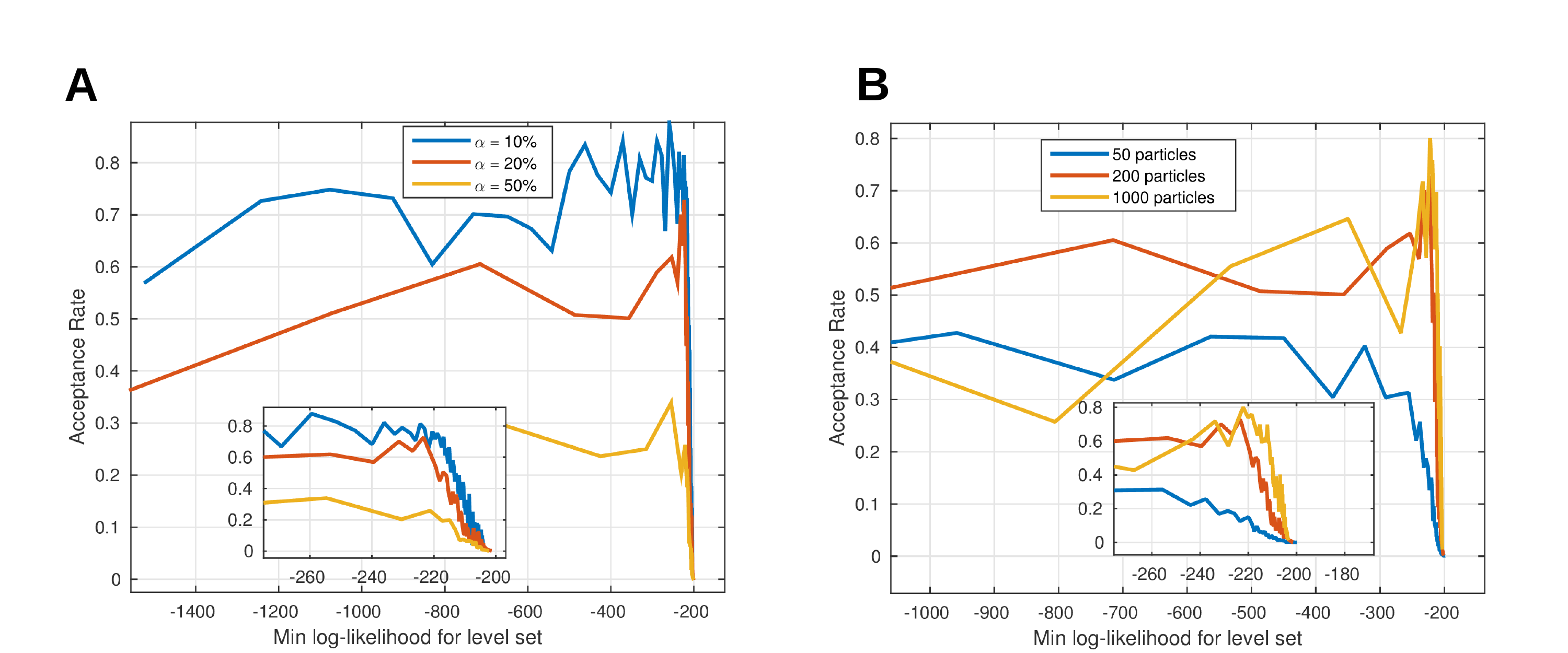}
\caption{{\bf A:} Acceptance rate for each log-likelihood threshold for runs with different values for $\alpha$. {\bf B:} Acceptance rate for each log-likelihood threshold for runs with different numbers or particles for the likelihood approximation.}
\label{fig:lac_diff_acceptance}
\end{figure}

\end{itemize}

\section{Examples used}\label{sec:sup models}

\subsection{Egf-Ras-Raf-Mek-Erk model}\label{sec:erk_model}
The Egf-Ras-Raf-Mek pathway is a signalling pathway, activated by Egf at the membrane and signalling through Ras, Raf and Mek to Erk (see Figure \ref{fig:models} B for an illustration). We use the model introduced in  \cite{ryu2015frequency},  with a slow negative feedback on Erk through Dusp and a fast, Egf-receptor dependent, negative feedback through a modelling species named ``NFB''. 
We use initial conditions different than in \cite{ryu2015frequency}. The involved species and corresponding initial concentrations can be taken from Table \ref{tab:erk_species}. The fixed values were taken from the literature. The initial concentration of the active form of the proteins is assumed 0, since the cells were starved before the experiment. The other values were taken from \href{http://bionumbers.hms.harvard.edu/}{http://bionumbers.hms.harvard.edu/}. For Dusp it was assumed that at the beginning of the experiment the concentration of mRNA and Protein was the same. 

\begin{figure}
\centering
\includegraphics[width=\textwidth]{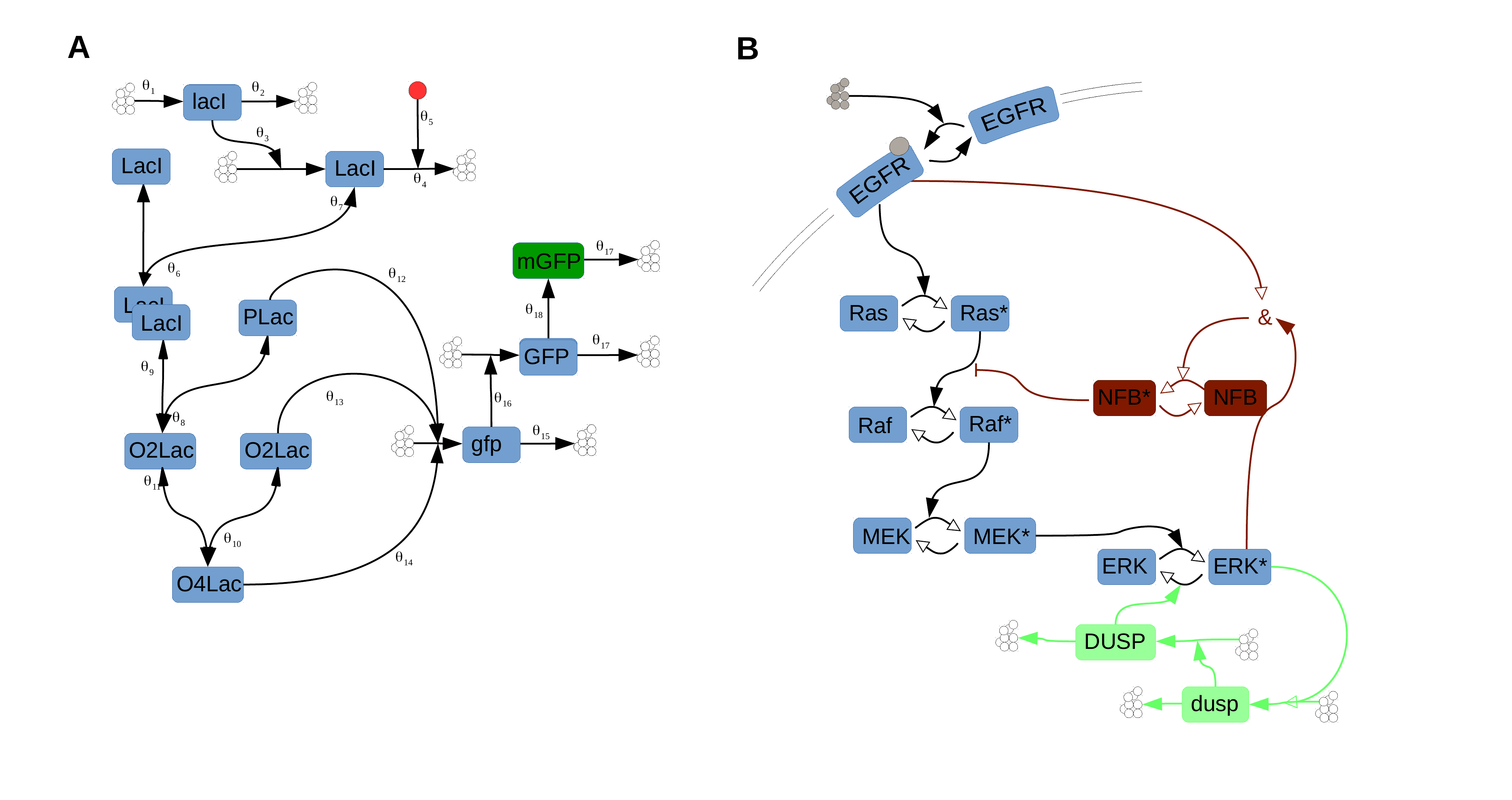}
\caption{{\bf A}: Schematic representation of the LacGfp model. {\bf B}: Schematic representation of the Egf-Ras-Raf-Mek-Erk model. }
\label{fig:models}
\end{figure}

\begin{table}[tb]
\label{tab:erk_species}
\caption{Species and intial concentration of the Ras-Raf-MEK-ERK model.}
\begin{tabular}{l l c }

\hline \hline

Species & Notation 
& Initial value prior interval / fixed value  \\

\hline

Egf Receptor & EGFR 
& [$\exp(-8)$, $\exp(1)$] \\

Active Egf Receptor & EGFR$^*$
& fixed to 0\\

Ras concentration &Ras
& fixed to 0.1\\

Active Ras concentration& Ras$^*$
& fixed to 0 \\

Raf concentration& Raf
&fixed to 0.7 (b-Raf: 0.2, c-Raf: 0.5) \\

Active Raf concentration& Raf$^*$
& fixed to 0 \\

MEK concentration& MEK
&fixed to 0.68 \\

Active MEK concentration& MEK$^*$
& fixed to 0 \\

ERK concentration& ERK
&fixed to 0.26 \\

Active ERK concentration& ERK$^*$
& fixed to 0 \\

NFB concentration& NFB
& [$\exp(-8)$, $\exp(1)$] \\

Active NFB concentration & NFB$^*$
& fixed to 0\\

Dusp mRNA/ Dusp Protein concentration& dusp/DUSP
& [$\exp(-8)$, $\exp(1)$] \\
\hline

\hline

\end{tabular}

\end{table}

The reaction equations take the following form:

\subsubsection{Reaction Equations} \label{sec:egf_reactions}
\begin{small}
$$ \dot{\textnormal{EGFR}}(t) = - (k_{1}\textnormal{egf}_{input}(t)\textnormal{EGFR}(t) )+(\gamma_{1} {\textnormal{EGFR}^*(t)}) $$
$$ \dot{\textnormal{EGFR}}^*(t) =(k_{1}\textnormal{egf}_{input}(t)\textnormal{EGFR}(t) ) -(\gamma_{1} {\textnormal{EGFR}^*(t)}) $$
$$\dot{\textnormal{Ras}}(t) = -\left(k_{2}\textnormal{EGFR}^*(t)\frac{\textnormal{Ras}(t)}{K_2+\textnormal{Ras}(t)}\right)+\left(\gamma_2\frac{\textnormal{Ras}^*(t)}{D_2+\textnormal{Ras}^*(t)}\right)$$ 
$$\dot{\textnormal{Ras}}^*(t) = \left(k_{2}\textnormal{EGFR}^*(t)\frac{\textnormal{Ras}(t)}{K_2+\textnormal{Ras}(t)}\right)-\left(\gamma_2\frac{\textnormal{Ras}^*(t)}{D_2+\textnormal{Ras}^*(t)}\right)$$
$$\dot{\textnormal{Raf}}(t) = - \left(k_3\textnormal{Ras}^*(t)\frac{\textnormal{Raf}(t)}{K_3+\textnormal{Raf}(t)}\frac{\textnormal{K}_{NFB}^2}{\textnormal{K}_{NFB}^2 + {\textnormal{NFB}^*(t)}^2}\right)+\left(\gamma_3\frac{\textnormal{Raf}^*(t)}{D_3+\textnormal{Raf}^*}\right) $$
$$\dot{\textnormal{Raf}}^*(t) =  \left(k_3\textnormal{Ras}^*(t)\frac{\textnormal{Raf}(t)}{K_3+\textnormal{Raf}(t)}\frac{\textnormal{K}_{NFB}^2}{\textnormal{K}_{NFB}^2 + {\textnormal{NFB}^*(t)}^2}\right)-\left(\gamma_3\frac{\textnormal{Raf}^*(t)}{D_3+\textnormal{Raf}^*}\right) $$
$$\dot{\textnormal{MEK}}(t) = -\left(k_4 \textnormal{Raf}^*(t)\frac{\textnormal{MEK}(t)}{K_4+\textnormal{MEK}(t)}\right)+\left(\gamma_4 \frac{\textnormal{MEK}^*(t)}{D_4+\textnormal{MEK}^*(t)}\right)$$ 
$$\dot{\textnormal{MEK}}^*(t) = \left(k_4 \textnormal{Raf}^*(t)\frac{\textnormal{MEK}(t)}{K_4+\textnormal{MEK}(t)}\right)-\left(\gamma_4 \frac{\textnormal{MEK}^*(t)}{D_4+\textnormal{MEK}^*(t)}\right)$$ 
$$\dot{\textnormal{ERK}}(t) = -\left(k_5 \textnormal{MEK}^*(t) \frac{\textnormal{ERK}(t)}{K_5+\textnormal{ERK}(t)}\right)+\left( \gamma_5 \textnormal{DUSP}(t) \frac{\textnormal{ERK}^*(t)}{D_5+\textnormal{ERK}^*}\right)$$ 
$$\dot{\textnormal{ERK}}^*(t) =\left(k_5 \textnormal{MEK}^*(t) \frac{\textnormal{ERK}(t)}{K_5+\textnormal{ERK}(t)}\right)-\left( \gamma_5 \textnormal{DUSP}(t) \frac{\textnormal{ERK}^*(t)}{D_5+\textnormal{ERK}^*}\right)$$ 
$$\dot{\textnormal{NFB}}(t) = -\left(k_6 \textnormal{ERK}^*(t)\frac{\textnormal{NFB}(t)}{K_6+\textnormal{NFB}(t)}\frac{{\textnormal{EGFR}^*(t)}^2}{K_{6R}^2+{\textnormal{EGFR}^*(t)}^2}\right)+\left(\gamma_6 \frac{\textnormal{NFB}^*(t)}{D_6+\textnormal{NFB}^*(t)}\right) $$
$$\dot{\textnormal{NFB}}^*(t) =\left(k_6 \textnormal{ERK}^*(t)\frac{\textnormal{NFB}(t)}{K_6+\textnormal{NFB}(t)}\frac{{\textnormal{EGFR}^*(t)}^2}{K_{6R}^2+{\textnormal{EGFR}^*(t)}^2}\right)-\left(\gamma_6 \frac{\textnormal{NFB}^*(t)}{D_6+\textnormal{NFB}^*(t)}\right) $$
$$ \dot{\textnormal{dusp}}(t) = \left(\textnormal{dusp}_{basal}\left(1+\textnormal{dusp}_{ind}\frac{{\textnormal{ERK}^*}^2}{K_{dusp}+{\textnormal{ERK}^*}^2}\right)\frac{ \textnormal{log}_{10}(2)}{\textnormal{T}_{dusp}}\right)-\left(\textnormal{dusp}(t)\frac{\textnormal{log}_{10}(2)}{\textnormal{T}_{dusp}}\right) $$
$$\dot{\textnormal{DUSP}}(t) = \left(\textnormal{dusp}(t)\frac{\textnormal{log}_{10}(2)}{\textnormal{T}_{DUSP}}\right)-\left(\textnormal{DUSP}(t)\frac{\textnormal{log}_{10}(2)}{\textnormal{T}_{DUSP}}\right) $$
\end{small}

\begin{table}[tb]
\centering

\caption{Prior distributions of the Ras-Raf-MEK-ERK parameters.}
\begin{tabular}{c l c c }

\hline \hline

Parameter & Meaning
& Prior interval & $\bar{\bm{\theta}}$ \\

\hline

$k_1$ & Egf-dependent Egf-Receptor activation rate
& [$\exp(-10)$, $\exp(5)$] \\

$\gamma_1$ & Egf-Receptor deactivation rate
& [$\exp(-10)$, $\exp(5)$] \\

$k_2$ & Egf-Receptor dependent Ras activation rate
& [$\exp(-10)$, $\exp(5)$] \\

$K_2$ & Hill constant for Ras activation
& [$\exp(-10)$, $\exp(5)$] \\

$\gamma_2$ & Ras deactivation rate
& [$\exp(-10)$, $\exp(5)$] \\

$D_2$ &Hill constant for Ras deactivation
& [$\exp(-10)$, $\exp(5)$] \\

$k_3$ & Ras dependent Raf activation rate
& [$\exp(-10)$, $\exp(5)$] \\

$K_3$ & Hill constant for Raf activation
& [$\exp(-10)$, $\exp(5)$] \\

$K_{NFB}$ & Coupleing constant for negative feedbock of NFB on Raf activation
& [$\exp(-10)$, $\exp(5)$] \\

$\gamma_3$ & Raf deactivation rate
& [$\exp(-10)$, $\exp(5)$] \\

$D_3$ & Hill constant for Raf deactivation
& [$\exp(-10)$, $\exp(5)$] \\

$k_4$ & Raf dependent MEK activation rate
& [$\exp(-10)$, $\exp(5)$] \\

$K_4$ & Hill constant for MEK activation
& [$\exp(-10)$, $\exp(5)$] \\

$\gamma_4$ & MEK deactivation rate
& [$\exp(-10)$, $\exp(5)$] \\

$D_4$ &Hill constant for MEK deactivation
& [$\exp(-10)$, $\exp(5)$] \\

$k_5$ & MEK dependent ERK activation rate
& [$\exp(-10)$, $\exp(5)$] \\

$K_5$ & Hill constant for ERK activation
& [$\exp(-10)$, $\exp(5)$] \\

$\gamma_5$ & DUSP dependent ERK deactivation rate
& [$\exp(-10)$, $\exp(5)$] \\

$D_5$ &Hill constant for ERK deactivation
& [$\exp(-10)$, $\exp(5)$] \\

$k_6$ & ERK and Egf-Receptor  dependent NFB activation rate
& [$\exp(-10)$, $\exp(5)$] \\

$K_6$ & Hill constant for NFB dependent NFB activation
& [$\exp(-10)$, $\exp(5)$] \\

$K_{6R}$ & Hill constant for Egf-Receptor dependent NFB activation
& [$\exp(-10)$, $\exp(5)$] \\

$\gamma_6$ & NFB deactivation rate
& [$\exp(-10)$, $\exp(5)$] \\

$D_6$ &Hill constant for NFB deactivation
& [$\exp(-10)$, $\exp(5)$] \\

$\textnormal{dusp}_{basal}$ & Basal dusp transcription rate
& [$\exp(-10)$, $\exp(5)$] \\

$\textnormal{dusp}_{basal}$ & ERK induced dusp transcription rate
& [$\exp(-10)$, $\exp(5)$] \\

$K_{dusp}$ & Hill constant for ERK dependent dusp transcription
& [$\exp(-10)$, $\exp(5)$] \\

$T_{dusp}$ & Half life of dusp
& fixed to 90 min \\

$T_{DUSP}$ & Half life of DUSP
& fixed to 90 min\\

$\textnormal{egf}_{input}(t)$ & Egf input at time $t$
& input parameter\\
\hline

\hline

\end{tabular}

\end{table}

\subsubsection{Noise Model}
The measurements are obtained using a FRET sensor (see \cite{ryu2015frequency}) and then normalized, such that the average of the signal is 1 at the beginning of the experiment. 
FRET sensors consist of a sensing unit, that has two fluorescent proteins on it. This sensing unit responds to particular biochemical activity with conformational changes that either change the distance of the two fluorescent proteins of the sensor, or their orientation. In both cases the result is that stimulation with light in the donor channel stimulated the donor protein, which, when excited, partially also stimulates the acceptor protein. Thus upon stimulation in the donor channel one is also able to measure a response in the acceptor channel. This conformational changes in the FRET protein are reversible, such  that an increase in the measured acceptor channel upon donor stimulation is proportional to the desired signal to be measured. The observed signal is taken to be the ration of the acceptor response and the donor response upon donor stimulation. 
For modelling we followed a similar approach as in \cite{birtwistle2011linear}. We took the final measurement to be 
$$ R(t) = \lambda \left(\frac{\textnormal{ERK}^*(t)E f_{AA}}{(\textnormal{ERK}(t) +(1-E) \textnormal{ERK}^*(t))f_{DD}} + \textnormal{bg}\right) + \epsilon_{\sigma},$$
where $\epsilon_{\sigma}$ is a zero mean normal random variable with $\sigma$ standard deviation. All parameters and their priors can be found in table \ref{tab:fret}.
\begin{table}[tb]
\centering
\label{tab:fret}
\caption{Prior distributions of the FRET Measurement model.}
\begin{tabular}{c l c }

\hline \hline

Parameter & Meaning
& Prior interval  \\

\hline

$\textnormal{bg}$ & background fluorescence
& [$\exp(-7)$, $\exp(6)$] \\

$\lambda$ & Scaling constant
& [$\exp(-7)$, $\exp(6)$] \\

$\sigma$ & Standard deviation of the normal noise on the measurement
& [$\exp(-8)$, $\exp(-3)$] \\

$E$ &\begin{tabular}{l}FRET efficiency (energy transferred from stimulated donor molecule\\to acceptor molecule in active FRET probes)\end{tabular} 
& [$\exp(-4)$, $\exp(-0.5)$] \\

$f_{AA}$ & Fraction of acceptor emission captured by the acceptor channel
& [$\exp(-8)$, $\exp(5)$] \\

$f_{DD}$ & Fraction of donor emission captured by the donor channel
& [$\exp(-8)$, $\exp(5)$] \\

\hline
\hline

\end{tabular}
\end{table}

\subsubsection{Inferred Posterior} 
We ran the LNS algorithm for the Egf-Ras-Raf-Mek-Erk model using trajectories corresponding to 6 differnt input patterns. The chosen trajectories are the means of responses to those different input patterns. The algorithm was run with $\alpha = 50\%$, 50 DP-GMM iterations for the density approximations. The inference was done on the Brutus Cluster of the ETH Zurich (\url{https://www1.ethz.ch/id/services/list/comp_zentral/cluster/index_EN}). Figure \ref{fig:erk_post_param} shows the posteriors obtained for the parameters of the model. 
 Figure \ref{fig:erk_post_initial} shows the posterior for the initial concentrations of the Egf Receptors, the model species NFB and of Dusp mRNA and protein. Figure \ref{fig:erk_post_fret} shows the posterior for the parameters of the FRET model. It can be seen that several parameters are very well identified. Since we don't know the actual value of those parameters, the only way to see if those parameters are reasonable is to look at the resulting simulations (Main paper Figure 3 A and B). The simulations seem to reproduce the behaviour of the system faithfully. To validate if the chosen model is appropriate we could now  compare the inferred posterior with measured biological quantities (for instance the initial concentration of Dusp). We can also draw conclusions for the model from the reaction parameters. For instance, the marginal posterior for $\textnormal{K}_{NFB}$ is concentrated at around 0.001, which is very low, compared to other parameters. Taking a look at the reaction equations \ref{sec:egf_reactions}, we see that such a low value for  $\textnormal{K}_{NFB}$ means that the model response to almost any increase in  $\textnormal{NFB}^*$ in a switch like manor. Similarly, a very large value for $\textnormal{D}_6$ indicates that the deactivation of $\textnormal{NFB}^*$ does not seem to play any role. These are just two illustrations what kind of conclusions can be drawn from an available posterior over the parameters and in what way the posterior may aid in further developing an available model.

%\begin{figure}
%\centering
%\includegraphics[width=\textwidth]{erk_param_post.eps}
%\caption{Posterior for the model parameters for the Egf-Ras-Raf-Mek-Erk model.}
%\label{fig:erk_post_param}
%\end{figure}
%
%
%\begin{figure}
%\centering
%\includegraphics[width=0.7\textwidth]{erk_initial_post.eps}
%\caption{Posterior for the inferred initial concentrations for the Egf-Ras-Raf-Mek-Erk model.}
%\label{fig:erk_post_initial}
%\end{figure}
%
%\begin{figure}
%\centering
%\includegraphics[width=0.7\textwidth]{erk_fret_post.eps}
%\caption{Posterior for the FRET noise model parameters.}
%\label{fig:erk_post_fret}
%\end{figure}
%

\subsection{LacGfp example}\label{sec:Lac_example}
The second model we use for the demonstration of our algorithm is rather large. It is for a LacGfp system with 9 species and 18 reactions. A schematic representation of the system is given in Figure \ref{fig:models} A. Table \ref{tab:lac_species} shows the species involved and their initial distribution for the simulation.

\begin{table}[tb]
\label{tab:lac_species}
\caption{Species and intial numbers of the LacGfp model}
\begin{tabular}{l l c }

\hline \hline

Species & Notation %& Unit 
& Initial Distribution\\

\hline

LacI mRNA & $lacI$ 
&$\textnormal{U}([0, 5])$ \\

LacI protein monomer & $LACI$
&$\textnormal{U}([0, 10])$\\

LacI dimer &$LACI2$
& fixed to 0\\

Unoccupied (active) Lac promoter & $PLac$
& fixed to 0 \\

Occupied Lac promoter with 2 repressor molecules bound& $O2Lac$
&fixed to 0 \\

Occupied Lac promoter with 4 repressor molecules bound&$O4Lac$
&$\textnormal{U}([50, 70])$ \\

GFP mRNA & $gfp$
&fixed to 0 \\

 ``Dark'' GFP protein& $GFP$
& fixed to 0 \\

Mature GFP protein& $mGFP$
&fixed to 0 \\

\hline

\hline
\end{tabular}

\begin{flushleft}
 $\textnormal{U}([a, b])$ denotes the uniform distribution on the interval $[a, b]$. 
\end{flushleft}

\end{table}

The reactions of the model all follow mass action kinetics and take the following form: \\

\subsubsection{Reactions}

\begin{enumerate}
\begin{small}
\item $ \emptyset \stackrel{\theta_{1}}{\longrightarrow} lacI $. Transcription of lacI mRNA (constitutive).
\item $ lacI \stackrel{\theta_{2}}{\longrightarrow} \emptyset $. Degradation of lacI mRNA (constitutive).
\item $ lacI \stackrel{\theta_{3}}{\longrightarrow} lacI + LACI $. Translation of LACI protein.
\item $ LACI \stackrel{\theta_u}{\longrightarrow} \emptyset $, where $\theta_u = \theta_4 + \theta_5 [IPTG]$. Degradation of LACI protein, increased by the input (IPTG).
\item $ LACI + LACI \stackrel{\theta_{6}}{\longrightarrow} LACI2 $. Dimerization of LACI protein.
\item $ LACI2 \stackrel{\theta_{7}}{\longrightarrow} LACI + LACI $. Dissociation of LACI dimer.
\item $ LACI2 + PLac \stackrel{\theta_{8}}{\longrightarrow} O2Lac $. Binding of LACI dimer to Lac operator sequence.
\item $ O2Lac \stackrel{\theta_{9}}{\longrightarrow} LACI2 + PLac $. Dissociation of LACI dimer from operator sequence.
\item $ O2Lac + O2Lac \stackrel{\theta_{10}}{\longrightarrow} O4Lac $. Binding of two LacI/operator complexes and tetramerization.
\item $ O4Lac \stackrel{\theta_{11}}{\longrightarrow} O2Lac + O2Lac $. Dissociation of tetramer structure.
\item $ PLac \stackrel{\theta_{12}}{\longrightarrow} PLac + gfp $. Transcription of gfp mRNA from active Lac promoter.
\item $ O2Lac \stackrel{\theta_{13}}{\longrightarrow} O2Lac + gfp $. Transcription of gfp mRNA from Lac promoter bound to LacI dimer.
\item $ O4Lac \stackrel{\theta_{14}}{\longrightarrow} O4Lac + gfp $. Transcription of gfp mRNA from Lac promoter bound to LacI tetramer.
\item $ gfp \stackrel{\theta_{15}}{\longrightarrow} \emptyset $. Degradation of gfp mRNA.
\item $ gfp \stackrel{\theta_{16}}{\longrightarrow} gfp + GFP $. Translation of dark GFP protein.
\item $ GFP \stackrel{\theta_{17}}{\longrightarrow} \emptyset $. Degradation of dark GFP protein.
\item $ GFP \stackrel{\theta_{18}}{\longrightarrow} mGFP $. Maturation of GFP.
\item $ mGFP \stackrel{\theta_{17}}{\longrightarrow} \emptyset $. Degradation of mature GFP protein.
\end{small}
\end{enumerate}

The effect of IPTG on the system is modelled as an increase in the degradation rate of $LACI$. The relationship between such rate and the inducer concentration, denoted $[IPTG]$, is assumed to be linear. We take the $[IPTG]$ concentration to be 10 $\mu M$. There is only $P=1$ measured species, namely $mGFP$. \\

The parameters used to simulate the dataset $\bm{y}$ as well as the priors used for the inference are shown in Table \ref{tab:priors}.

\begin{table*}[tb]
\centering

\caption{Prior distributions of the Lac-GFP parameters and the real values $\bm{\theta}^*$ used for the simulation of $\bm{y}$.}
\begin{tabular}{c l c c c}

\hline \hline

Parameter & Meaning
& Prior interval & $\bm{\theta}^*$ \\

\hline

$\theta_1$ & lacI transcription rate
& [1, 50] & 1.215\\

$\theta_2$ & lacI degradation rate
& [0.1, 30] & 7.430\\

$\theta_3$ & LACI translation rate
& [1, 50] &  1.439\\

$\theta_4$ & IPTG-independent LACI degradation rate
& [0.1, 20] & 3.971\\

$\theta_5$ & IPTG-induced increase in LACI degradation rate
& [1, 100] & 1.173\\

$\theta_6$ & Dimerization rate of LACI
& [0.16, 3000] & 1716.683\\

$\theta_7$ & Dissociation rate of LACI dimers
& [0.1, 20]  &  6.282\\

$\theta_8$ & Binding rate of LACI dimers to Lac promoter
& [0.0016, 1.6] & 0.433\\

$\theta_9$ & Dissociation rate of LACI dimers from Lac promoter
& [0.1, 20] & 0.679\\

$\theta_{10}$ & Tetramerization rate of LACI
& [0.16, 1.600] & 173.576\\

$\theta_{11}$ & Dissociation rate of LACI tetramers
& [0.1, 20] &0.623\\

$\theta_{12}$ & gfp transcription rate from free PLac promoter
& [1, 150] & 119.232\\

$\theta_{13}$ & gfp transcription rate from LACI dimer-bound PLac
& [0.1, 15] &0.120 \\

$\theta_{14}$ & gfp transcription rate from LACI tetramer-bound PLac
& [10$^{-5}$, 1] &0.006\\

$\theta_{15}$ & gfp degradation rate
& [0.1, 20] & 1.673\\

$\theta_{16}$ & GFP translation rate
& [1, 50] & 32.318\\

$\theta_{17}$ & GFP degradation rate
 & [0.1, 20]  & 1.196\\

$\theta_{18}$ & GFP maturation rate
 & [0.1, 20] &2.102 \\

\hline

\hline

\end{tabular}
\begin{flushleft}

For all the Lac-Gfp inference problems presented in this paper, each parameter was assigned an independent uniform log prior distribution in the interval listed in the table.

\end{flushleft}

\label{tab:priors}

\end{table*}

We assume that our observation $y_\tau$ at a certain time $\tau$ is distributed according to 
$$y_\tau \sim \mathcal{N}(22.85  x_\tau, 5.72\sqrt{x_\tau}) + \mathcal{B},$$
where $x_\tau$ is the number of GFP molecules  at time $\tau$, $\mathcal{N}(\mu, \sigma)$ is the normal distribution with mean $\mu$ and standard deviation $\sigma$ and $\mathcal{B}$ is a known background fluorescence actually taken from flow cytometry data. The distribution of the background fluorescence is shown in Figure \ref{fig:first 10 traj Lac Gfp} A. This noise model implies a mean fluorescence of $22.85$ for each protein and a standard deviation of $5.72$. 
Figure \ref{fig:first 10 traj Lac Gfp} B shows the first 10 simulated trajectories of the Lac-Gfp system measured on 29 timepoints. As can be seen, most trajectories exhibit switch like  behaviour. This is indeed one of the properties of the considered example that make it particularly hard to perform likelihood approximation for the single trajectories. 

\begin{figure}
\centering
\includegraphics[width=\textwidth]{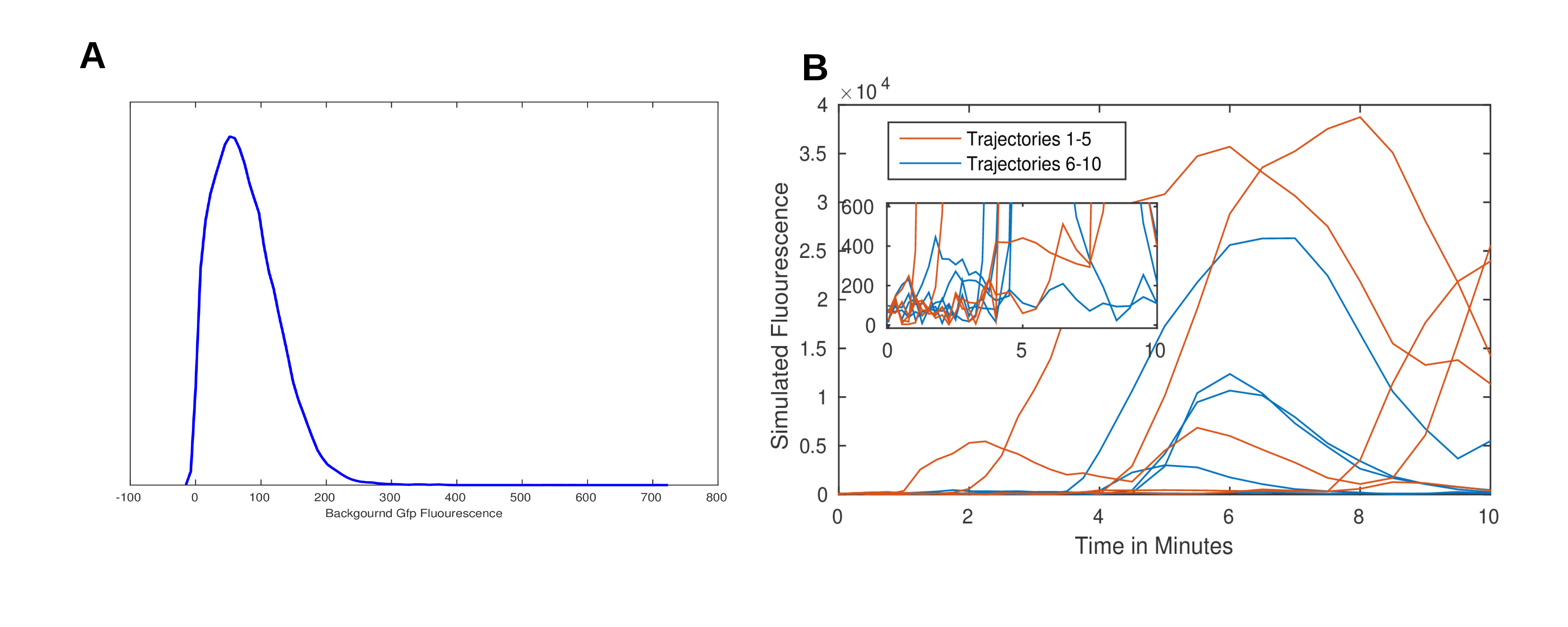}
\caption{{\bf A}: The distribution of the background fluorescence. {\bf B}: 10 trajectories of simulated Lac-Gfp data, measured at 29 time points. }
\label{fig:first 10 traj Lac Gfp}
\end{figure}

\section{Tuning the algorithm} \label{sec:lns params}

\subsection{Different values for $\alpha$}\label{sec:sup diff alpha}
We discuss the choice of $\alpha$ by running the algorithm with  different values for the quantile $\alpha$. The number of particles used for the approximation of the likelihood was 200 and only one observed trajectory was considered. We performed the parameter inference with $\alpha = 10\%, 20\%$ and $50\%$. The corresponding developments of the minimal and maximal log-likelihood can be seen in Figure \ref{fig:lac_diff_alpha} B. As expected we can see that choosing $\alpha$ small ($10\%$) results in a slow increase of log-likelihoods, while a higher $\alpha$ corresponds to a steeper increase. The minimal and maximal log-likelihood seems to converge to the same value for each different values of $\alpha$. Figure Figure \ref{fig:lac_diff_alpha} A shows the marginals of the posterior obtained using different values of $\alpha$. We see that, for the most part, each run seems to approximate the same posterior. The minor differences (for instance in the marginal for $\theta_2$ can be explained with a slightly different exploration of the parameter space. 

\begin{figure}
\centering
\includegraphics[width=\textwidth]{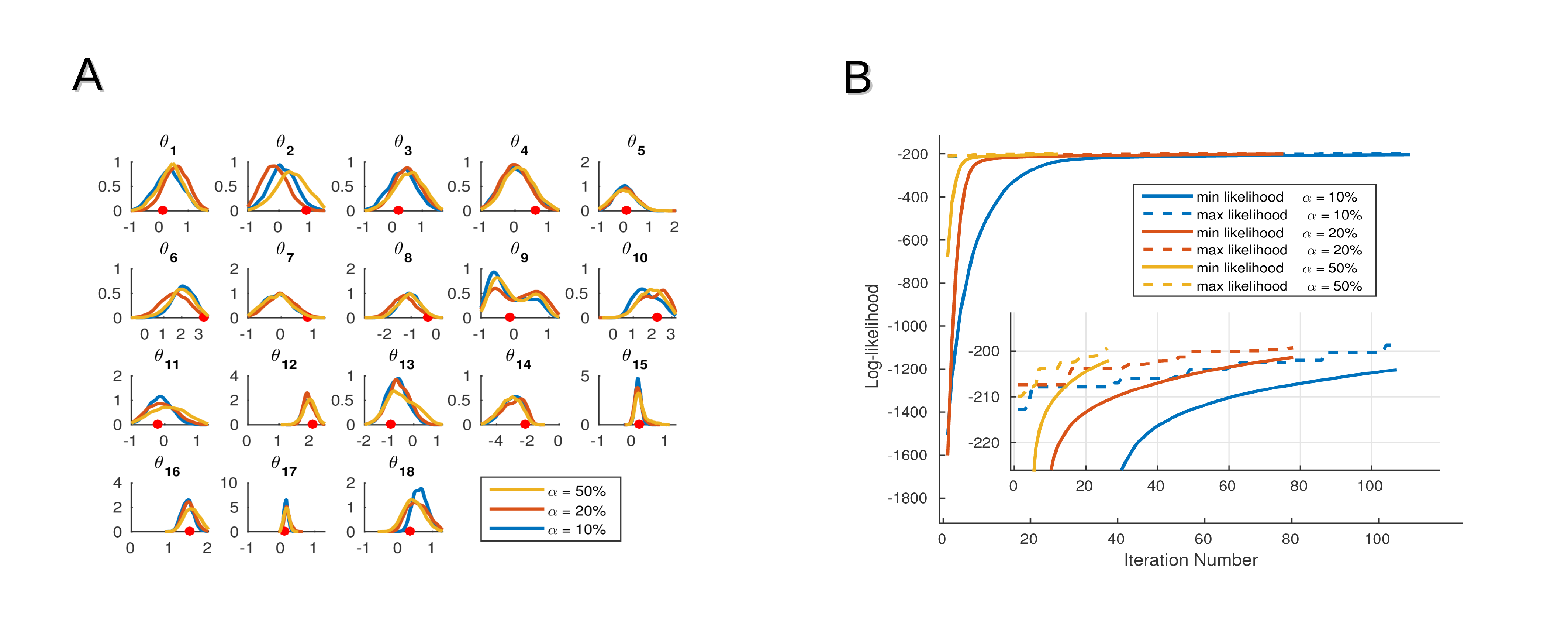}
\caption{{\bf A:} Marginal posteriors obtained for the Lac-Gfp model using different values for $\alpha$. {\bf B:} Minimal and maximal log-likelihood for each level set, for different values for $\alpha$.}
\label{fig:lac_diff_alpha}
\end{figure}

\subsection{Accuracy of likelihood computation}
The approximation of the likelihood is what decides whether a particle is accepted or not, thus the accuracy of the likelihood approximation is expected to play an important role for the inference of the posterior. We applied our algorithm to the Lac-Gfp model, using only one observed trajectory and different numbers of particles for the likelihood approximation. The resulting posteriors are shown in Figure \ref{fig:lac_diff_post} A.

\begin{figure}
\centering
\includegraphics[width=\textwidth]{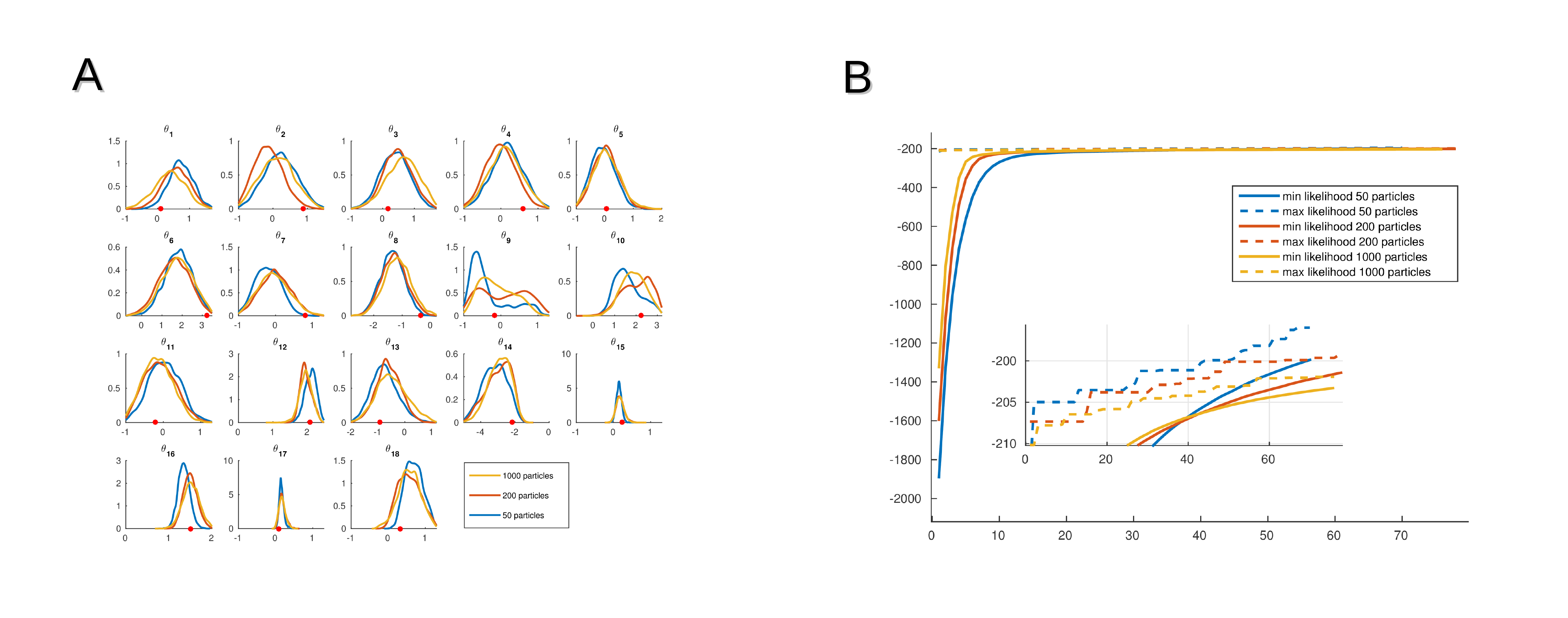}
\caption{{\bf A:} Marginal posteriors obtained for the Lac-Gfp model using different number particles for the likelihood approximation. A higher number of particles indicates a more accurate posterior approximation. {\bf B:} Minimal and maximal log-likelihood for each level set, for different particle number for the likelihood approximation.}
\label{fig:lac_diff_post}
\end{figure}

As can be seen, for the most part the posteriors do not seem to differ too much. Only for parameter $\theta_9$, the posterior obtained with only 50 particles seems to differ significantly from the one obtained using 200 and 1000 particles. However, 50 particles are indeed very little and considering the switching behaviour of the system, it is not surprising that the estimated posterior differs from the posterior obtained using 200 or 1000 particles for the likelihood approximation. However, the difference between 200 and 1000 particles for likelihood approximation seems to be very small indicating, that 200 particles are already enough to approximate the posterior accurately. Figure  \ref{fig:lac_diff_post} B illustrates the development of the minimal and maximal log-likelihood for each level set for different number of particles for the likelihood approximation. It can be seen that for all particle numbers the gap between minimal and maximal log-likelihood gets smaller. As the likelihood approximation gets more precise (the number of particles increases), the minimal and maximal log-likelihoods tend to get smaller, when compared to less precise approximations.

%\bibliography{bibliography}

\end{document}